\documentclass[structabstract]{aa} 

\usepackage[dvips]{graphicx}
\usepackage{amssymb}
\usepackage{xcolor} 
\usepackage{soul} 
\usepackage[normalem]{ulem}
\usepackage{float}
\usepackage{ragged2e}
\usepackage{caption}
\usepackage{subcaption}
\usepackage{psfrag}
\usepackage{graphicx}
\usepackage{txfonts}
\usepackage{atbegshi}
\AtBeginDocument{\AtBeginShipoutNext{\AtBeginShipoutDiscard}}
\usepackage{natbib}
\bibpunct{(}{)}{;}{a}{}{,}
\def\kms{{\rm km\, s^{-1}}}
\begin{document}
   \titlerunning{Overdensity of VVV galaxies behind the Galactic bulge}
   \authorrunning{Galdeano et al.}
   \title{Overdensity of VVV galaxies behind the Galactic bulge}.
%  \subtitle{}
   \author{Daniela Galdeano\inst{1}, 
          Luis Pereyra\inst{2}, 
          Fernanda Duplancic\inst{1},
          Georgina Coldwell\inst{1},
          Sol Alonso\inst{1},
          Andr\'es N. Ruiz\inst{2,3},
          Sof\'ia A. Cora\inst{4,5},
          Noelia Perez\inst{1},
          Cristian Vega-Mart\'inez\inst{6,7},
          \and Dante Minniti\inst{8,9,10}
          }
   \institute{Departamento de Geof\'{i}sica y Astronom\'{i}a, CONICET, Facultad de Ciencias Exactas, F\'{i}sicas y Naturales, Universidad Nacional de San Juan, Av. Ignacio de la Roza 590 (O), J5402DCS, Rivadavia, San Juan, Argentina             \and  
    Instituto de Astronom\'ia Te\'orica y Experimental, CONICET-UNC, Laprida 854, X5000BGR, C\'ordoba, Argentina
    \and
    Observatorio Astron\'omico de C\'ordoba, UNC, Laprida 854, X5000BGR, C\'ordoba, Argentina
    \and 
    Instituto de Astrof\'isica de La Plata, CONICET-UNLP, Observatorio Astron\'omico, Paseo del Bosque s/n, B1900FWA, La Plata, Argentina
    \and
    Facultad de Ciencias Astron\'omicas y Geof\'isicas, UNLP, Observatorio Astron´omico, Paseo del Bosque s/n, B1900FWA, La Plata, Argentina
    \and
    Instituto de Investigaci\'on Multidisciplinar en Ciencia y Tecnolog\'ia, Universidad de La Serena, Ra\'ul Bitr\'an 1305, La Serena, Chile
    \and
    Departamento de Astronom\'ia, Universidad de La Serena, Av. Juan Cisternas 1200 Norte, La Serena, Chile
    \and
    Departamento de F\'isica, Facultad de Ciencias Exactas, Universidad Andres Bello, Av. Fernandez Concha 700, Las Condes, Santiago, Chile
    \and
    Millennium Institute of Astrophysics, Av. Vicuna Mackenna 4860, 782-0436, Santiago, Chile
    \and
    Vatican Observatory, V00120 Vatican City State, Italy}
             
   \date{Received xxx; accepted xxx}

   \abstract
% context heading (optional)
   {The extragalactic vision we have through the Milky Way is very unclear. There is significant extinction of the optical emission from objects located in the region called the Zone of Avoidance (ZOA). NIR wavelengths are less affected by extinction, and therefore the infrared surveys in this zone are a potential source of astronomical discoveries. Nevertheless, these observations need to be compared with cosmological simulations in order to carry out high$-$accuracy studies.
}
% aims heading (mandatory)
   {Our aim is to identify extragalactic sources in the ZOA, using infrared images of the VVV survey. We consider mock galaxy catalogues in order to interpret observational results.
}
% methods heading (mandatory)
   {We studied a region of 1.636 square degrees corresponding to the VVV tile $b204$. Using SExtractor, we analysed photometric data generating a catalogue of extended sources in this area. In order to confirm these sources as galaxy candidates we visually inspected RGB images looking for typical galaxy features. Using 2MASX and GCMW catalogued sources we tested completeness and contamination of our catalogue and define suitable colour cuts to select galaxies. 
   We also compared the observational results with those obtained from two semi-analytical models on Dark Matter simulations. One galaxy catalogue was constructed with the SAG semi-analytic model of galaxy formation, and the other one was constructed with the L-Galaxies semi-analytic model. 
}
% results heading (mandatory)
   {By adopting CLASS\_STAR$< 0.5$, $r_{1/2} > 0.7$ arcsec and specific colour cuts (J-Ks$>$0.97, J-H$>$0 and H-Ks$>$0) we generated an automatic catalogue of extended sources. After visual inspection we identified 624 sources with 10$<$Ks$<$17 as galaxy candidates. The contamination of the automatic catalogue is 28\% when considering visually confirmed galaxies as reliable objects. The estimated completeness is 87\% up to magnitude Ks=13.5. We analysed the spatial distribution of galaxy candidates, finding a high concentration of galaxies in a small region of 15 arcmin radius. This region has three times higher density than similar areas in the tile. We compared the number of galaxies in this small area with the mean density values obtained from a suitable sample of galaxies from semi-analytic models finding that our results are consistent with an overdensity region. 
}
% Conclusions heading (mandatory)
  {Using VVV near-infrared data and mock catalogues we detect new extragalactic sources that have not been identified by other catalogues. We demonstrate the potentiality of the VVV survey in finding and studying a large number of galaxy candidates and extragalactic structures obscured by the Milky Way.}
 
\maketitle 

%
%________________________________________________________________
\section{Introduction}
\label{intro}

The discovery of new extragalactic astronomical objects
is a topic of interest in the  mapping of large$-$scale structures and the  distribution  of  mass  in  the universe. Therefore, several large astronomical surveys have been developed with the aim being to detect new sources at different wavelengths (e.g Sloan Digital Sky Survey-SDSS, \cite{blanton17}; Faint Images of the Radio Sky at Twenty-Centimeters-FIRST, \cite{bec95}; Two Micron All Sky
Survey-2MASS, \cite{skr06}; etc.) covering a significant fraction of the sky. However, 
the area obscured by the Milky Way leads to an incomplete picture of large$-$scale structures; the dust and stars therein obstruct optical observations. This region, called the Zone of Avoidance (ZOA), poses an observational challenge.

To meet this challenge, several galaxy catalogues have been developed. Optical catalogues \citep{kra00, wou04} have allowed the detection of new galaxies at low Galactic latitude, however the observations are hampered and gathering of information is hampered by Galactic absorption. Furthermore, near-infrared (NIR), X-ray and H1 radio surveys \citep{Roman1998, ebe02, Vauglin2002, kor04, pat05, skr06, Huchra2012} have detected galaxies and galaxy clusters at low Galactic latitude.

During the last two decades, enormous efforts has been made to reveal the extragalactic structures behind the ZOA. 
For example \cite{Jarrett2000} developed basic algorithms and operations toward detection, identification, characterisation, and extraction of extended sources from the Two Micron All-Sky Survey (2MASS) catalogue. Also, \cite{nag04} presented deep near-infrared survey with 19 obvious galaxies and 38 galaxy candidates in a 36 x 36 arcmin$^{2}$ region centred on the giant elliptical radio galaxy PKS 1343–601, which was found to be the core of an unknown rich cluster located in the Great Attractor region and \cite{ske09} published a deep Ks band photometric catalogue with 390 sources (235 galaxies and 155 galaxy candidates) around the core of the rich nearby Norma cluster (ACO3627), in a region of 45 x 45 arcmin$^{2}$.

More recently, using the HI Parkes All-Sky Survey, \cite{sta16} identified new structures in the Great Attractor region, although there is still a huge area that has not been explored.
Moreover, \cite{schr19} published a catalogue with 170 galaxies located in the northern region of the ZOA, extracted from the shallow version of the blind HI survey with the Effelsberg 100m radio telescope, EBHIS. These large-scale structures could be part of possible filaments at the edge of the local volume.
In addition, \cite{mac19} presented redshifts for 1041 2MASS Redshift Survey galaxies that previously lacked this information and updated measurements for another 27.

In particular, the near infrared public survey VISTA Variables in V\'ia L\'actea;\citep{min10,sai12} (VVV) has proven to be an excellent tool finding and studying extragalactic objects in the ZOA due to its depth ($\sim$ three magnitudes deeper than 2MASS) and high angular resolution.
For example, \cite{amo12} identified 204 new galaxy candidates from the VVV photometry of 1.636 square degrees near the Galactic plane, increasing the surface density of known galaxies behind the Milky Way by more than an order of magnitude. Furthermore, \cite{col14} found the VVV NIR galaxy counterparts of a new cluster of galaxies at redshift $\mathrm{z} = 0.13$ detected in X-ray by SUZAKU \citep{mor13}. 
In addition, \cite{bar18} found 530 new galaxy candidates in two different tiles in the region of the Galactic disk. These latter authors detected and characterised the candidates using a combination of SExtractor and PSFEx techniques. Recently, \cite{bar19} confirmed the existence of these galaxy candidates with spectroscopic data using Flamingos-2 at GEMINI Observatory. All these works confirm the potential of the VVV for detecting new extragalactic objects.

In recent years, different extragalactic works have focused on comparing results from observations and simulations. Semi-analytic models (SAMs) of galaxy formation have been designed to reproduce the physical properties of observed galaxy distributions in cosmological volumes. These models use the merger trees of dark matter haloes extracted from numerical dark-matter-only simulations as a backbone and, using physical motivated recipes, populate those haloes with semi-analytic galaxies, describing their formation and evolution \citep[see, for instance,][and references therein]{som15,kne18}.

The primary goal of this project is to analyse VVV data in order to identify extragalactic sources. Moreover, we use mock catalogues to statistically predict the properties and counts of galaxies to provide an approximate estimate of the total number of galaxies hidden behind the Milky Way, (referred to as background galaxies hereafter) per VVV tile. The results from observations and simulations are compared in order to find overdensity regions in the VVV survey. 
This research represents a pilot test in a bounded region, such as a tile, with the aim of exploring the efficiency of the method. In the future, this technique could be automatised and applied to larger regions of the VVV and VVVX surveys.

This paper is structured as follows: in \S \ref{sec:vvv} we describe the VVV survey. Afterwards, in \S \ref{sec:galax} we explain the method to obtain the automatic catalogues with software SExtractor and the correspondent procedure used to analyse the VVV images. We present our general results in \S \ref{sec:results}. 
The SAMs and construction of a mock background galaxy catalogue and the results obtained from them are described in \S \ref{sec:mock}. In \S \ref{sec:Overdensity} we discuss the existence of an overdensity region in tile $b204$. Finally, we summarise our main conclusions in \S \ref{sec:conc}.
The adopted cosmology throughout this paper is $\Omega = 0.3$, $\Omega_{\Lambda} = 0.7$, and $H_0 = 100~ \kms \rm Mpc$.

\section{The VVV survey}
\label{sec:vvv}

The VVV survey is a NIR ESO (European Southern Observatory ) public survey setup to map the Milky Way bulge ($-10^o < l < +10^o$ and $-10^o < b < +5^o$) and disk ($ -65^o < l < -10^o$
and $-2.25^o < b < +2.25^o$) areas close to its centre obtained using the 4.1-m ESO VISTA (Visual and Infrared Survey Telescope for Astronomy; Emerson et al. 2004, 2006; Emerson \& Sutherland 2010) telescope, located in the Atacama Desert of northern Chile on Cerro Paranal at 2635m altitude. VISTA was built by a consortium of 18 universities from the United Kingdom. The telescope is 4 m class with an
altitude-azimuth mount, and quasi-Ritchey-Chretien optics. The instrument includes a wide-field corrector lens system, autoguider and active optics sensors.
Observations were carried out with the VISTA IR CAMera (VIRCAM), which is composed by 16 Raytheon VIRGO 2048x2048 HgCdTe on CdZnTe substrate science detectors, with a mean pixel scale of 0.339 arcsec px$^{-1}$ and a field of view per exposure of 0.59 square degrees. The detectors are arranged in a grid with spaces in between detectors of 90$\%$ of the detector width in the x-direction and 42.5$\%$ in the y-direction. Six exposures are required to make a contiguous area of 1.636 square degrees. The individual exposure is known as a 'pawprint' and the final area as a 'tile'. 

The VVV area covers ~562 deg$^{2}$, with 196 tiles needed to map the bulge and 152 tiles for the disk, and is fully imaged in five photometric bands: Z, Y, J, H, and Ks. Furthermore, the survey has up  to  80  multi-epoch Ks observations  spread  over  approximately  7-8  years,  for  variability  and  proper  motion  studies.  More  details  can  be  found  in \cite{min10} and \cite{sai12}. 
The observations were processed within the  VISTA Data  Flow System (VDFS) pipeline at  the  Cambridge Astronomical Survey Unit (CASU\footnote{http://casu.ast.cam.ac.uk/surveys-projects/vista}; \cite{lewis2010}). The VDFS provides object catalogues automatically generated for each target frame with astrometric and photometric calibration information into FITS binary tables. Standard object descriptors are used for morphological object classification, such as aperture flux measurements, intensity-weighted centroid estimates, and shape information including half$-$light radius and class star parameter.

In the present work, we use data from the tile $b204$ located in the Galactic bulge at $l=355.182^o$ and $b=-9.68974^o$ ($\alpha=273.72^o$ and $\delta=-37.87^o$).
This tile was selected because of its position at the edge of the bulge, which implies a low stellar density and also low extinction, relative to the tiles in the center of the bulge and disc, within the VVV survey area.
Searching for extragalactic sources in the tile $b204$ allows us to test a reliable method for galaxy identification in the bulge area of the VVV survey.

 In addition, the tile $b204$ represents a region where 13 galaxies have already been catalogued by 2MASX\footnote{ \scriptsize 2MASXJ18123243-3759580, 2MASXJ18130421-3740314, 2MASXJ18131374-3743504, 2MASXJ18131474-3740354, 2MASXJ18132074-3739236, 2MASXJ18144659-3739470, 2MASXJ18144727-3744140, 2MASXJ18145268-3716080, 2MASXJ18154360-3838524, 2MASXJ18161057-3729587, 2MASXJ18164121-3816133, 2MASXJ18164505-3747263, 2MASXJ18172135-3721158}\citep{Jarrett2000} and five galaxy candidates were discovered by \cite{Saito1990}\footnote{\scriptsize CGMW4-00457, CGMW4-00298, CGMW4-00340, CGMW4-00408, CGMW4-00415}. The results obtained from this research can therefore also be compared with a very well-known sample of bright extended sources.

\section{Source detection}
\label{sec:galax}

The VVV survey was optimised to detect stars in the Milky Way, but nevertheless the automatic catalogues provide a flag for non-stellar objects \citep[CS=+1,][]{sai12}. For the tile $b204$ we download CASU catalogues finding that about 70000 sources have this flag category in J, H and Ks bands, and therefore it is expected that this parameter is unreliable for the identification of extended sources as galaxy candidates. In the present work, we create a new catalogue from the VVV images, using SExtractor \citep{ber96}, because it is better suited to treating extended sources. To this end we download J, H and Ks images of the tile $b204$ from CASU .

In this section, we describe the techniques used to identify sources in the observational data from the VVV survey. 

\subsection{Automatic detection}
\label{automatic}

The SExtractor software was developed to detect, deblend, measure and classify sources from astronomical images. Several steps are followed to identify and separate objects from the background noise, where different parameters should be settled (see \cite{ber96} for more details). Moreover, crowded fields are very common in the VVV sky area and the deblending advantage from SExtractor is a useful feature to analyse overlapping objects.

In order to generate a multi-wavelength catalogue for the analysis, we used SExtractor in double-image mode with the Ks image as a reference because of its better quality. 
A Ks image is used for source detection, and the second, H or J-band image for measurements only. It is necessary that all images have identical size and be perfectly aligned. The number of detected sources will not affect their positional or basic shape parameters if we change the H or J images.
With this procedure it is not necessary to make a correlation between objects of different filters because all of them are in the same position.

In this way, we generated a multi-band catalogue with all detected sources. SExtractor recognised and separated from the background noise all the objects with a threshold more than twice the medium brightness of the sky and spanned over at least ten connected pixels on the VVV images \citep{col14,bar18}.
A filter convolution was applied over the images by SExtractor with the aim being to efficiently separate low-surface-brightness objects from spurious detections. For a proper detection of extended and faint sources we adopt a Gaussian filter, with 3 x 3 pixels with a convolution mask of a full width half maximum of three pixels. In addition, for the deblending process we used the default values, because, according to our tests, they allowed the overlapped stars to be efficiently separated from galaxies in this tile.

Following the procedure suggested by CASU, we used the zero point magnitude\footnote{http://casu.ast.cam.ac.uk/surveys-projects/vista/technical/catalogue-generation} 
 to calibrate the magnitudes estimated by SExtractor. 
The output parameters selected from SExtractor catalogues were equatorial coordinates, J, H, and Ks magnitudes (total and three-pixel-radius circular aperture), and for star-galaxy separation parameters, we used CLASS\_STAR and half$-$light radius ($r_{1/2}$). For the total magnitudes, we used the MAG$\_$AUTO, which is based on Kron´s algorithm \citep{kro80}. We calculated the colours with J, H and Ks magnitudes in three-pixel radius apertures to compare with colours obtained from CASU \citep{col14}.

\begin{figure}[htb]
	\includegraphics[width=\columnwidth]{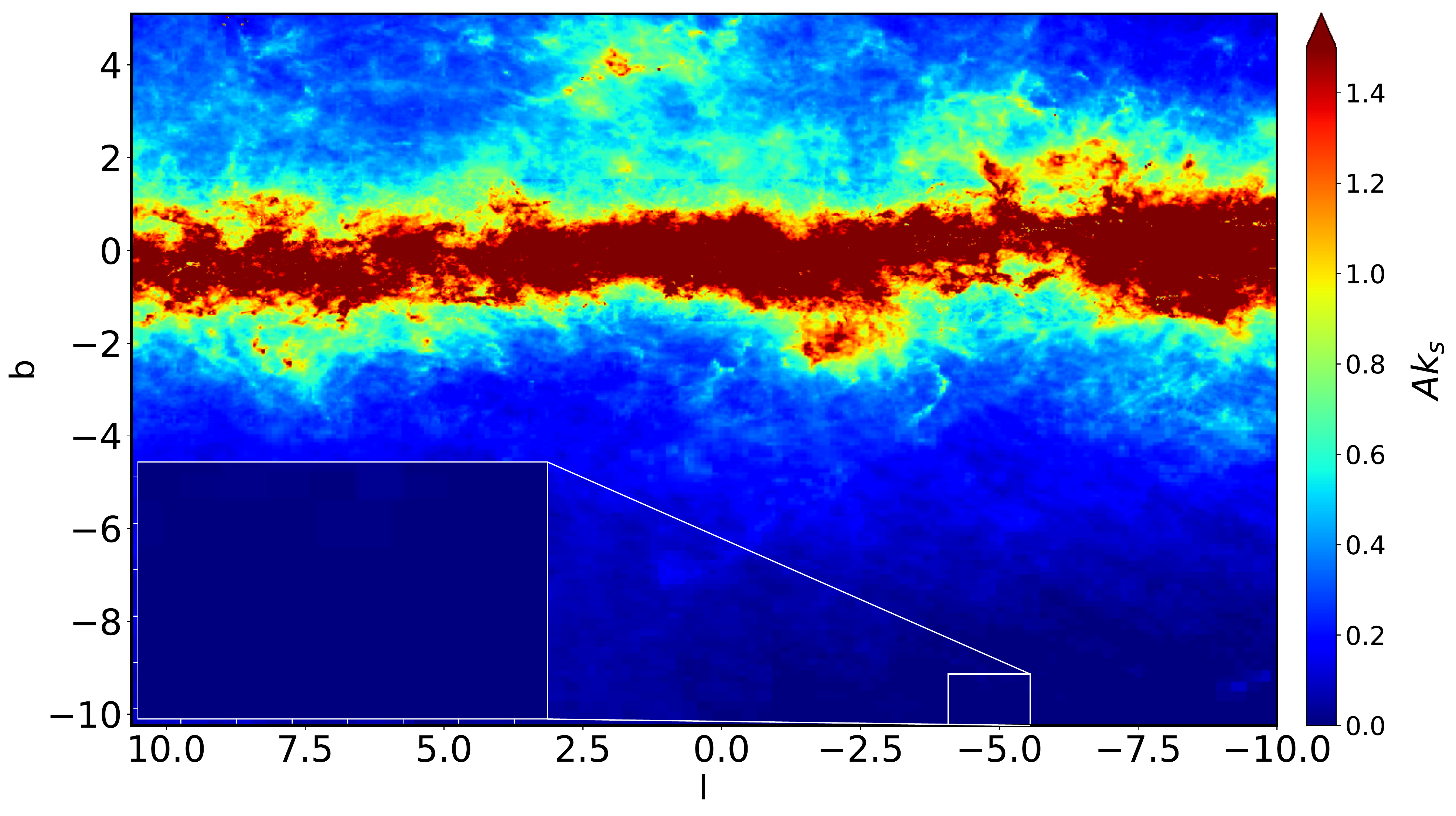}
    \caption{AKs 
    extinction map constructed for the VVV bulge region 
    using the data from \citet{gon12}.
    The colour scale is saturated for AKs values larger than $1.5$ mag. 
    A zoom-in to the tile $b204$ used in this work is outlined in the white colour box.
    }
    \label{fig:extinction_gonzalez}
\end{figure}

For the extinction correction we used the procedure described in \cite{gon12} adopting the \cite{nis09} extinction law because it is in good agreement with the measurements taken from \cite{gon12}, where $E(J-H) = 0.671 E(J-Ks)$. This high-resolution extinction map covers the complete region of the VVV survey, in particular the Galactic bulge. The map is available to the community in the web BEAM (Bulge Extinction And Metallicity) calculator\footnote{https://www.oagonzalez.net/beam-calculator}. This tool provides a lower limit for the mean extinction, AKs, and the colour excess, E(J-Ks), for a given set of coordinates and radius of the region. 

In Fig. \ref{fig:extinction_gonzalez} we show the extinction map constructed for the VVV bulge region using the data from \citet{gon12}.
From the extinction maps, the reddening in tile $b204$ is fairly low and uniform. The average values of the extinction corrections are 0.0051, 0.0028, and  0.0018 mag for J, H, and Ks filters, reaching maximum values of 0.041 in J, 0.023 in H, and 0.014 in Ks bands, respectively.
This can be seen clearly in the zoom-in of Fig. \ref{fig:extinction_gonzalez}.

Subsequently, considering all the mentioned issues, we obtained a total of 300976 sources with SExtractor, including stellar and extended objects. In Section \S \ref{sec:results}, we detail the procedure used to classify these sources taking into account the parameters obtained with the software. In order to make our final catalogue more solid, it was indispensable to verify the object classification and generate a visual catalogue. We describe our visual inspection in \S \ref{visual}.

\subsection{Photometric and astrometric comparison}

\begin{figure*}[htb]
\centering
    \includegraphics[width=0.94\textwidth]{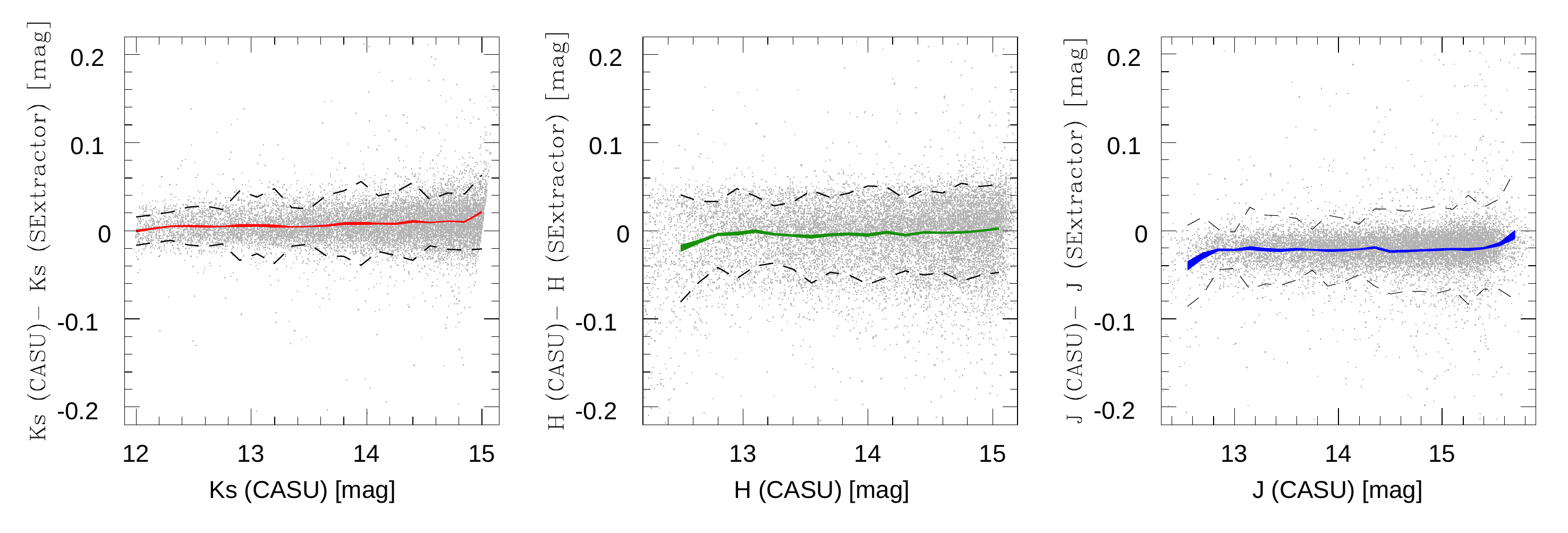}
        \caption{Differences in magnitudes between SExtractor and CASU stellar objects for filters J, H and Ks as a function of CASU magnitudes. Dashed lines show 1$\sigma$ boundaries and the thickness of the coloured line represents the standard error of the mean.}
        
    \label{calib}
\end{figure*}

In order to estimate the accuracy of our photometric and astrometric measurements we compared the SExtractor magnitudes and positions with those obtained from CASU using only point sources. 

The CASU catalogue provides a stellar classification through a flag indicating the most probable morphological classification.
To this end, we downloaded CASU data for tile $b204$ and constructed a VVV catalogue with stellar information in this region. Then we
selected sources, classified as stellar (CS=-1) that are fainter than Ks=12 to avoid saturated stars and brighter than Ks=15 which 
corresponds to the completeness limit for this tile.

On the other hand, from the automatic catalogue obtained using SExtractor, we used the CLASS\_STAR parameter to eliminate all objects with very high likelihood of being extended by adopting CLASS\_STAR $> 0.8$.
Moreover, in order to make a suitable comparison, we applied an aperture correction for the aperture magnitudes of these stellar objects by considering the apcor parameter provided by CASU$^4$.
To perform the comparison of magnitudes and positions between CASU and SExtractor catalogues, we identified common objects as those within a radius of 2 arcsec.

To estimate magnitude differences, we used the three-pixel-radius circular aperture magnitude for each filter in both catalogues. We obtained acceptable agreement between the VVV system and SExtractor magnitudes with an average and standard error of the mean of $\Delta Ks = 0.0071 \pm 0.0005, \Delta H = -0.0038 \pm 0.0018$ and $\Delta J=-0.0215 \pm 0.0017$, respectively. These results are shown in Fig. \ref{calib}.

Furthermore, we calculated the difference between SExtractor and CASU coordinates in order to check the position exactitude of our measurements. The SExtractor astrometry shows excellent agreement with the CASU data mean, reaching values of $\Delta\alpha=1.0 \times 10^{-4}$ and $\Delta\delta=0.5 \times 10^{-4}$ arcsec.

\section{Results}
\label{sec:results}

\subsection{Automatic catalogue of extended sources}

\begin{figure}[htb]
\centering
\includegraphics[width=90mm,height=80mm ]{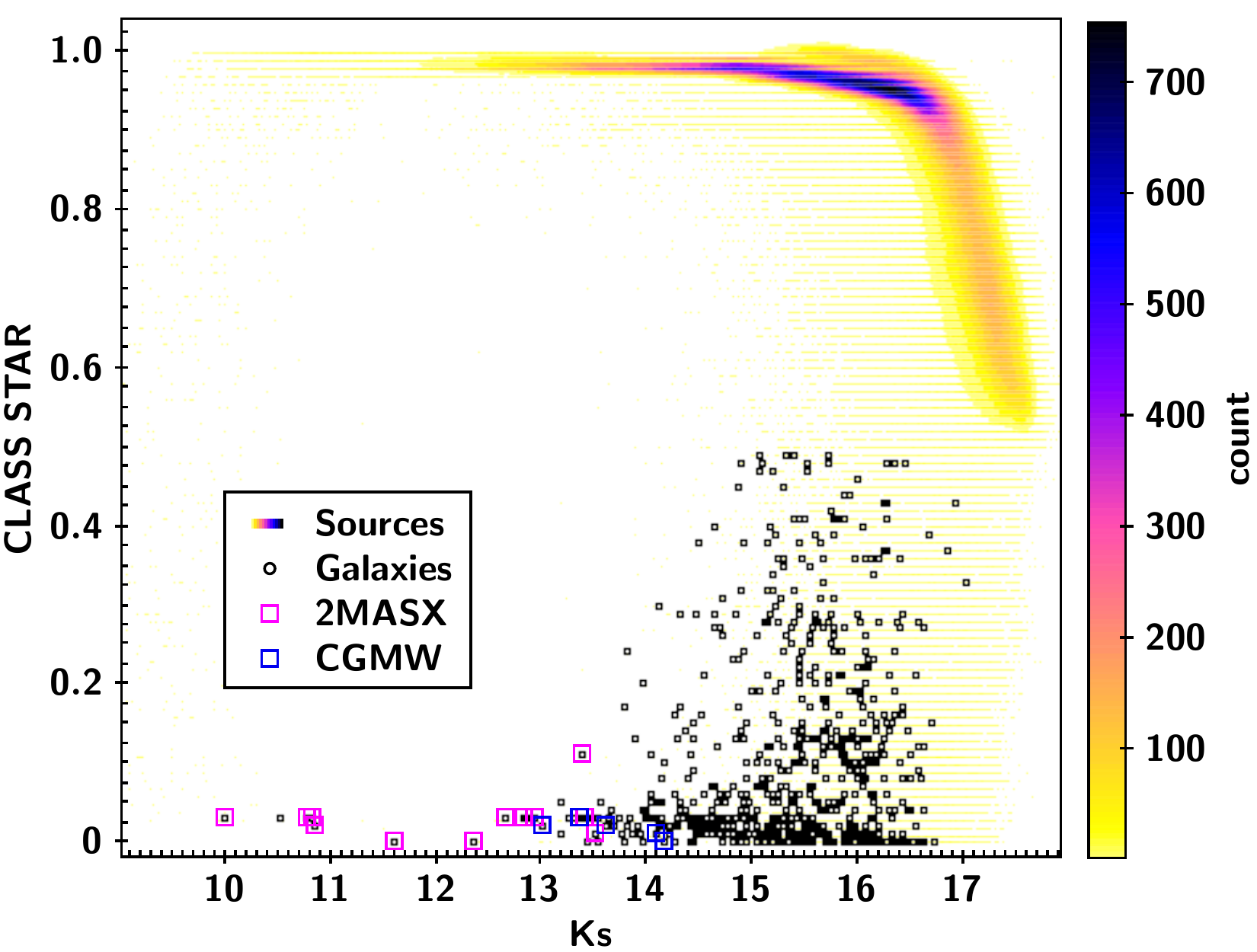}
\caption{SExtractor stellarity index CLASS\_STAR vs. the total Ks band source magnitude. The circles correspond to galaxy candidates. Galaxies catalogued by other surveys are shown as squares. The colour scale corresponds to number counts (bar on the right)}
\label{cKs}
\end{figure}

 In order to derive a catalogue of automatically selected extended sources, we consider different parameters. We used the star classification parameter CLASS\_STAR given by SExtractor. This parameter takes values of between 0 and 1 where the extended objects, with characteristics typical of galaxies, have values close to 0 and objects with values close to 1 have a high likelihood of being stars.
Figure~\ref{cKs} shows the relation between the CLASS\_STAR parameter and the total Ks magnitude for our dataset; in this figure and the successive ones the magnitudes and colours have been corrected by extinction. For the total magnitudes we used the MAG\_AUTO. The colours were obtained using the circular aperture magnitudes within a diameter of three pixels.
Taking into account the presence of two sequences, we removed star-like sources from our catalogue by adopting a conservative limit of CLASS\_STAR $<$ 0.5.

\begin{figure}[htb]
\centering
\includegraphics[width=90mm,height=80mm]{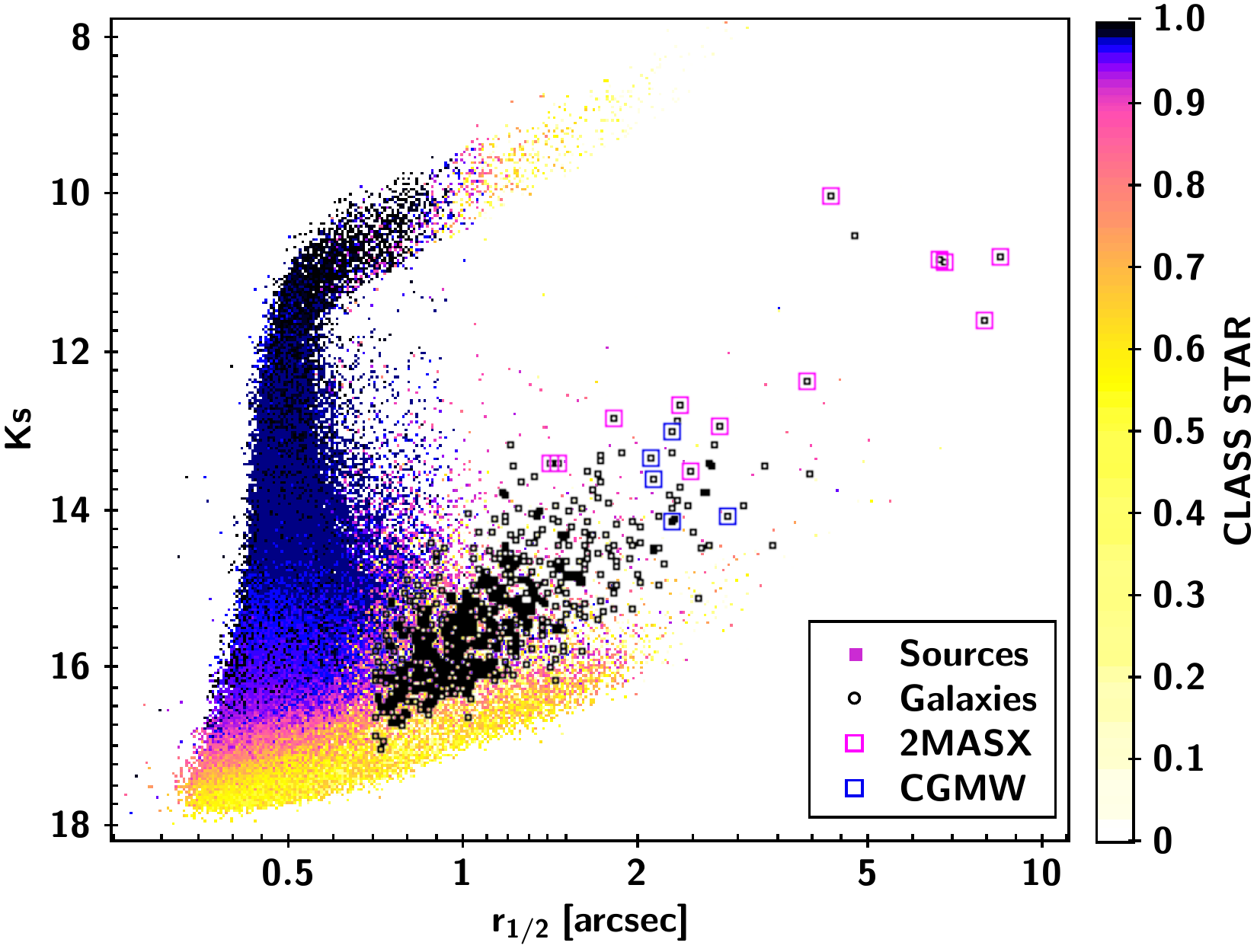}
\caption{Ks total magnitude vs. half$-$light size $r_{1/2}$ diagram for the whole sample of detected sources. The circles correspond to galaxy candidates. Galaxies catalogued by other surveys are shown in squares.
The points are colour-coded according to the SExtractor stellarity parameter CLASS\_STAR (bar on the right).}

\label{Ksflux}
\end{figure}

We also used another classification parameter in order to strengthen the separation between stars and galaxies. This parameter is the half$-$light radius ($r_{1/2}$), which encloses $50\%$ of the total flux and is a good parameter to complement the classification criteria \citep{col14}. Objects with CLASS\_STAR $\approx 0$ are expected to have values $r_{1/2} > 0.7$ arcsec because they are extended, reaching values of around 8 arcsec. 
In Fig.~\ref{Ksflux} we show the half$-$light radius versus total Ks magnitude
for the whole sample of sources detected in the tile b204. The points are colour-coded according to the stellarity parameter CLASS\_STAR (the bar on the right) showing that the majority of the objects with CLASS\_STAR $\approx$ 1 approximately populate the stellar locus at $r_{1/2}$ $<$ 0.7 arcsec, while sources with CLASS\_STAR $<$ 0.5 tend to have larger $r_{1/2}$.
Combining these two parameters, we establish as galaxy candidates those objects with a CLASS\_STAR $<$ 0.5 and $r_{1/2}$ $>$ 0.7 arcsec, obtaining a clean sample of 11144 sources.

Additionally, we consider different colour cuts to identify galaxy candidates. According to \cite{fin00} objects with J-Ks$<1$ are probably stellar objects with spectral types later than G5 and earlier than K5. Moreover, \citet{Jarrett2003} found that galaxies with different Hubble morphological types derived from 2MASS show colours with a mean J-Ks$\approx 1$ and H-Ks$\approx 0.27$.
Taking into account the star-galaxy separation, \citet{Lucas2008} found that the external galaxies form a well-defined region located at $1.5<$J-K$<3.0$, K$>14$, on the J-K colour-magnitude diagram, based on data from the UKIDSS Galactic plane survey.
Furthermore, when these authors included only defined galaxies from visual inspection, they considered a constraint of J-K$> 1.9$.
However, these magnitudes and colours have no extinction corrections applied.
Recently, \cite{schr19b} presented a homogeneous 2MASS bright galaxy catalogue at low Galactic latitudes ($|b| < $ $10.0^o$) and used an unobscured high-latitude sample to compare the dependence of the extinction correction.  
The authors found that, although the scatter in the colours is large since they are sensitive to extinction, the mean value is well defined: $<$ (J - Ks) $>$ $=$ 0.938 $\pm$ 0.002.
Nevertheless, the improved extinction correction changes the mean colour to 0.988 with a standard deviation of $\pm$0.12, which is in good agreement with the high-latitude sample.
 \cite {bar18} also define similar colour cuts to separate star from galaxy candidates. In addition, \cite {amo12} and \cite {col14} found that objects selected as galaxy candidates from the VVV survey show mainly J-Ks$> 1.0$ colours.
 
  Taking this into consideration, in the present work we considered  J-Ks$> 0.97$, J-H$> 0$ and H-Ks$>0$ to separate galaxy candidates from the stellar locus. These cuts agree well with an intrinsic galaxy colour of J-K$\approx 1.0$. The colour cuts adopted in this work were chosen to improve the reliability of  our catalogue. A detailed description of completeness and contamination analysis can be found in section \ref{contamination}.
 
 Using the constraints described above we obtain a final sample of 2439 automatically identified extended sources.

\subsection{Visual catalogue of galaxy candidates}
\label{visual}
\begin{figure*}[htb]
    \centering
    \includegraphics[width=0.99\textwidth]{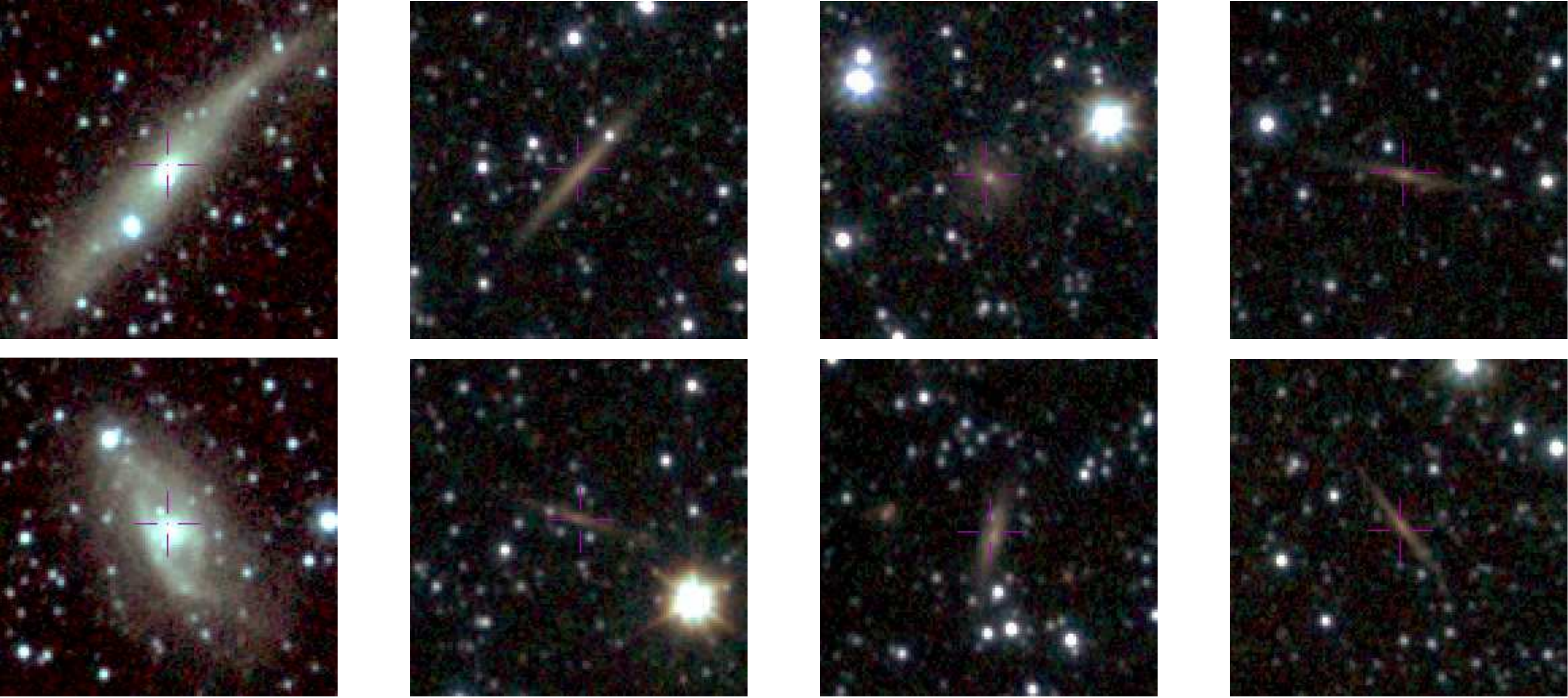}
    \caption{False-colour J(blue), H(green), and Ks(red) images of eight galaxy candidates detected in the tile b204, as examples. The two more-luminous galaxies in the left panels were also catalogued by 2MASX: 2MASX J18164505-3747263 (left-top) and 2MASX J18172135-3721158 (left-bottom). The length of each box side is 30 arcsec.}
    \label{galaxies}
\end{figure*}

Following \cite{amo12}, \cite{col14} and \cite{bar18}, we carried out a visual inspection in order to verify the automatic detection made with SExtractor. We performed a visual identification of galaxy candidates using Aladin \citep{bon00}. We constructed a false-colour RGB image with filters J, H and Ks. The objects with morphological features and surface brightness corresponding to galaxies were identified as visual galaxy candidates.
With this procedure, from the 2439 objects automatically classified as extended sources, 624 could be considered  as visual galaxies (hereafter galaxy candidates), with Ks magnitude in the range $10<$Ks$<17$. This is about $\approx 25\%$ of the sample of the automatic extended sources.  We show examples of these galaxy candidates in Fig.~\ref{galaxies}.

Table \ref{tab2} lists the main parameters of ten galaxy candidates ordered by Right Ascension. The list includes an identification
number, sexagesimal equatorial coordinates, three pixel aperture magnitudes, total magnitudes, CLASS\_STAR parameter and, half$-$light radius.

Extinction corrections have not been applied to the listed magnitudes, leaving the choice of the extinction map to the reader. The preferred extinction map can be obtained from different authors, for example \cite{gon12} and \cite{Schlafly2011}\footnote{https://irsa.ipac.caltech.edu/applications/DUST/}. The full table is available in electronic format.

\begin{table*}[htb]
\centering
\caption{Galaxy candidates catalogue}
\begin{tabular}{|c c c c c c c c c c c| }
\hline
ID & RA & Dec & Ks(3) & H(3) & J(3) & Ks & H & J & CLASS\_ & $r_{1/2}$ \\
 & (J2000) & (J2000) & (Aperture) & (Aperture) & (Aperture) & (Total) & (Total) & (Total) & STAR & [arcsec] \\

\hline
\hline

 01& 18:10:50.63 &-38:13:13.4 &  16.43$\pm$0.07 &  17.12$\pm$0.06 & 17.80$\pm$0.04 & 15.23$\pm$0.06 & 15.94$\pm$0.06 & 17.07$\pm$0.04 & 0.05 & 1.95 \\
 02& 18:10:54.29 &-38:07:00.7 &  13.90$\pm$0.02 &  14.23$\pm$0.01 & 14.90$\pm$0.01 & 13.17$\pm$0.01 & 13.46$\pm$0.01 & 14.05$\pm$0.01 & 0.05 & 1.21 \\
 03& 18:10:55.41 &-38:16:54.6 &  14.19$\pm$0.02 &  14.50$\pm$0.01 & 15.16$\pm$0.01 & 13.45$\pm$0.01 & 13.72$\pm$0.01 & 14.61$\pm$0.01 & 0.03 & 1.22 \\
 04& 18:10:59.23 &-38:13:00.8 &  15.88$\pm$0.05 &  16.64$\pm$0.05 & 17.61$\pm$0.04 & 15.57$\pm$0.05 & 16.26$\pm$0.04 & 16.85$\pm$0.03 & 0.07 & 0.80 \\
 05& 18:11:12.33 &-38:19:06.7 &  16.19$\pm$0.06 &  17.01$\pm$0.06 & 17.83$\pm$0.04 & 15.80$\pm$0.06 & 16.33$\pm$0.05 & 17.51$\pm$0.04 & 0.16 & 0.85 \\
 06& 18:11:13.22 &-38:04:33.4 &  16.65$\pm$0.08 &  17.49$\pm$0.08 & 18.30$\pm$0.06 & 16.28$\pm$0.07 & 17.04$\pm$0.07 & 17.65$\pm$0.05 & 0.01 & 0.89 \\
 07& 18:11:13.27 &-38:04:06.0 &  16.27$\pm$0.06 &  17.12$\pm$0.06 & 17.71$\pm$0.04 & 15.93$\pm$0.07 & 16.52$\pm$0.06 & 16.95$\pm$0.04 & 0.13 & 0.85 \\
 08& 18:11:14.48 &-38:04:32.5 &  16.13$\pm$0.05 &  16.78$\pm$0.05 & 17.55$\pm$0.03 & 15.92$\pm$0.06 & 16.45$\pm$0.05 & 16.67$\pm$0.04 & 0.44 & 0.71 \\
 09& 18:11:14.52 &-37:58:06.5 &  16.00$\pm$0.05 &  16.62$\pm$0.05 & 17.37$\pm$0.03 & 14.71$\pm$0.04 & 15.21$\pm$0.03 & 16.05$\pm$0.02 & 0.01 & 1.84 \\
 10& 18:11:18.36 &-38:05:41.2 &  16.29$\pm$0.06 &  16.96$\pm$0.05 & 17.82$\pm$0.04 & 15.74$\pm$0.07 & 16.50$\pm$0.06 & 17.44$\pm$0.04 & 0.04 & 1.00 \\
  
\hline  
\end{tabular}
\label{tab2}
\end{table*}

\subsection{Completeness and contamination analysis}
\label{contamination}

In order to test the reliability of our methods, we used 13 2MASX and five CGMW galaxies catalogued  in our field. We matched the equatorial coordinates of these objects with the sample of automatic extended sources detected by SExtractor, before the colour cut, considering a separation lower than 3*$r1/2$. Because most of the catalogued objects in b204 tile are 2MASX galaxies we considered the completeness limit of this sample \citep[Ks=13.5,][]{Jarrett2000} and restricted our analysis of completeness and contamination to the  magnitude range $10<$Ks$<13.5$. Under these constraints we have 15 catalogued sources (13 2MASX and two CGMW galaxies) to be compared with  76 automatically detected extended objects and 29 visually confirmed galaxy candidates in our field.

\begin{figure}
  \centering
  \includegraphics[width=\columnwidth]{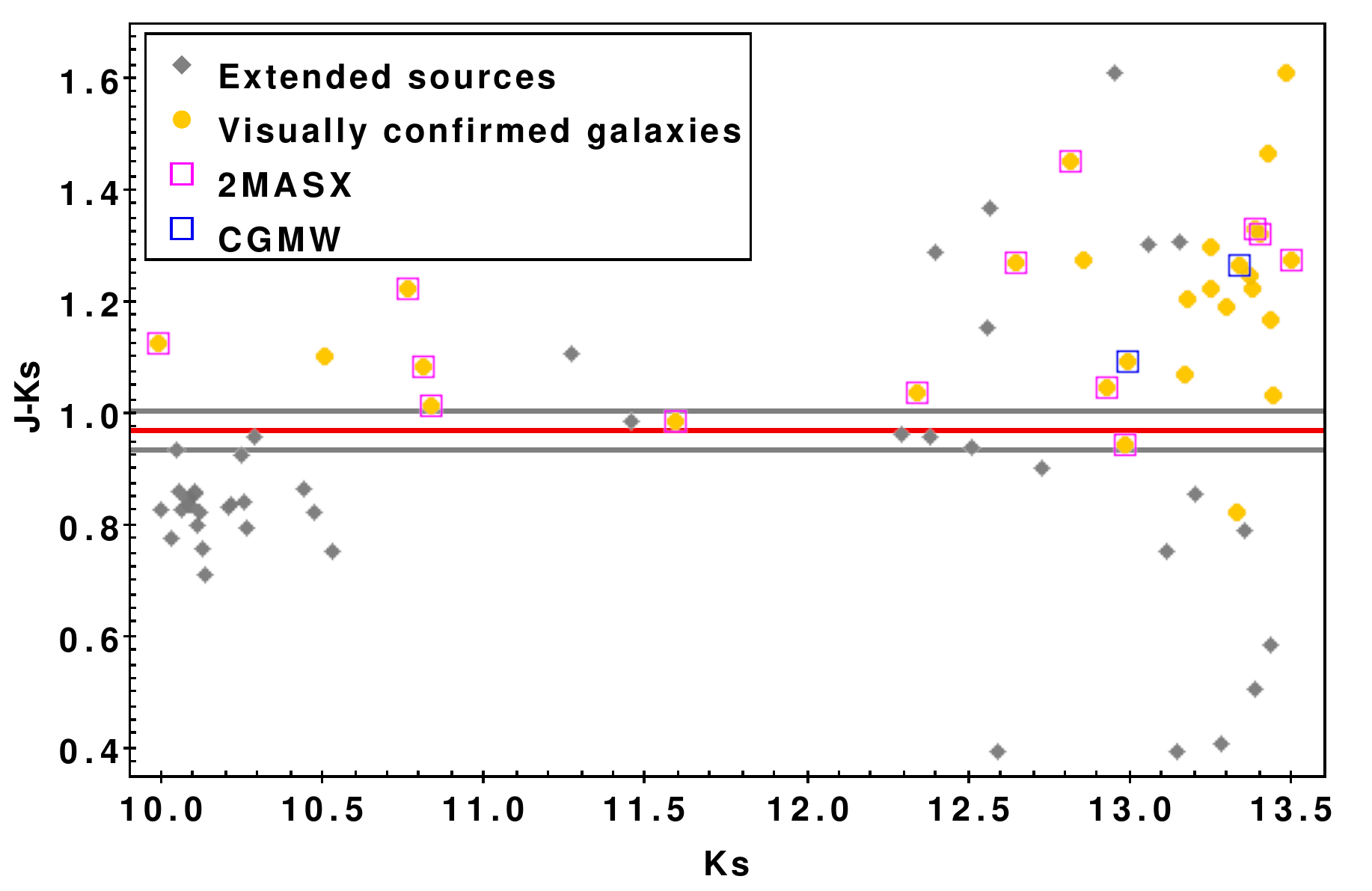}
\caption{Colour-magnitude diagram (J-Ks vs. Ks) of automatically detected extended sources, visually confirmed galaxy candidates, and catalogued sources in our field. The lines in this figure represent the different colour cuts used in the completeness and contamination analysis. The red line corresponds to the adopted limit J-Ks=0.97.} 
\label{CMag}
\end{figure}

Given that the adopted  J-H$>$0 and H-Ks$>$0 limits are conservative values, the completeness and contamination of our catalogues depend mostly on the J-Ks colour cut. Figure~\ref{CMag}shows a colour-magnitude diagram of automatically detected extended objects, visually confirmed galaxies, and catalogued sources. We tested different cuts based on the J-Ks colour of the three bluest 2MASX galaxies (the lines in Fig. \ref{CMag}). Table~\ref{tt} lists the adopted cuts and the number of objects in each catalogue. If we consider that only catalogued objects are reliable galaxies (an unrealistic assumption given that the VVV images are much deeper), by adopting a limit J-Ks=0.94 we include all 2MASX and CGMW sources, but the contamination of extended sources that have not been visually confirmed is 62\%.
For J-Ks=0.97 the automatic catalogue of extended sources has a contamination of 60\% and we identify 93\% of the catalogued galaxies.  
If we adopt a colour cut of J-Ks=1.0, the contamination of the extended sources catalogue is 60\% but completeness drops to 85\%. 
Moreover if now we consider visually confirmed galaxies as reliable objects, the contamination rates of the extended sources catalogue are 28\%, 23\%, and 21\% for the 0.94, 0.97, and 1.0 J-Ks limits, respectively. 
Therefore, we find a good compromise between completeness and contamination adopting J-Ks=0.97 (red line in Fig. \ref{CMag}).  Moreover, the election of this colour cut in the identification of galaxies using VVV data is reinforced from the Mock analysis (see Section \ref{sec:mock})

\begin{table}[htb]
\centering
\caption{Number of objects with 10$<$Ks$<$13.5 in the different samples used in the completeness and contamination tests.}
\begin{tabular}{|l c c c| }
\hline
Colour cut & Extended  & Visually confirmed  & Catalogued  \\
          & sources & galaxies & galaxies \\
\hline
\hline
J-Ks$>$0.94 & 39 & 28 & 15\\
J-Ks$>$0.97 & 35 & 27 & 14\\
J-Ks$>$1.0 & 33 & 26 & 13\\
J-Ks$<$0.97 & 37 & 2 & 1 \\
\hline  
\end{tabular}
\label{tt}
\end{table}

In addition to the above analysis, we  visually explored the RGB images of 37 SExtractor extended sources with 10$<$Ks$<$13.5 and J-Ks$<$0.97, finding 29 of these objects associated to bright stars (the bright extended sources in the bottom left region of Fig. \ref{CMag}) and six linked to image artefacts. Only two objects present galaxy features: one is an uncatalogued source with J-Ks=0.82 and the other is the 2MASXJ18131374-3743504 galaxy with J-Ks$=0.94$. Therefore, with our colour cuts we lose one 2MASX catalogue source and one extended object that was visually confirmed as a galaxy, and therefore the completeness of our visual catalogue can be estimated as 87\% (we identify 14 of 16 galaxies), up to magnitude Ks=13.5.

\begin{figure}
  \centering
 \includegraphics[width=0.95\columnwidth]{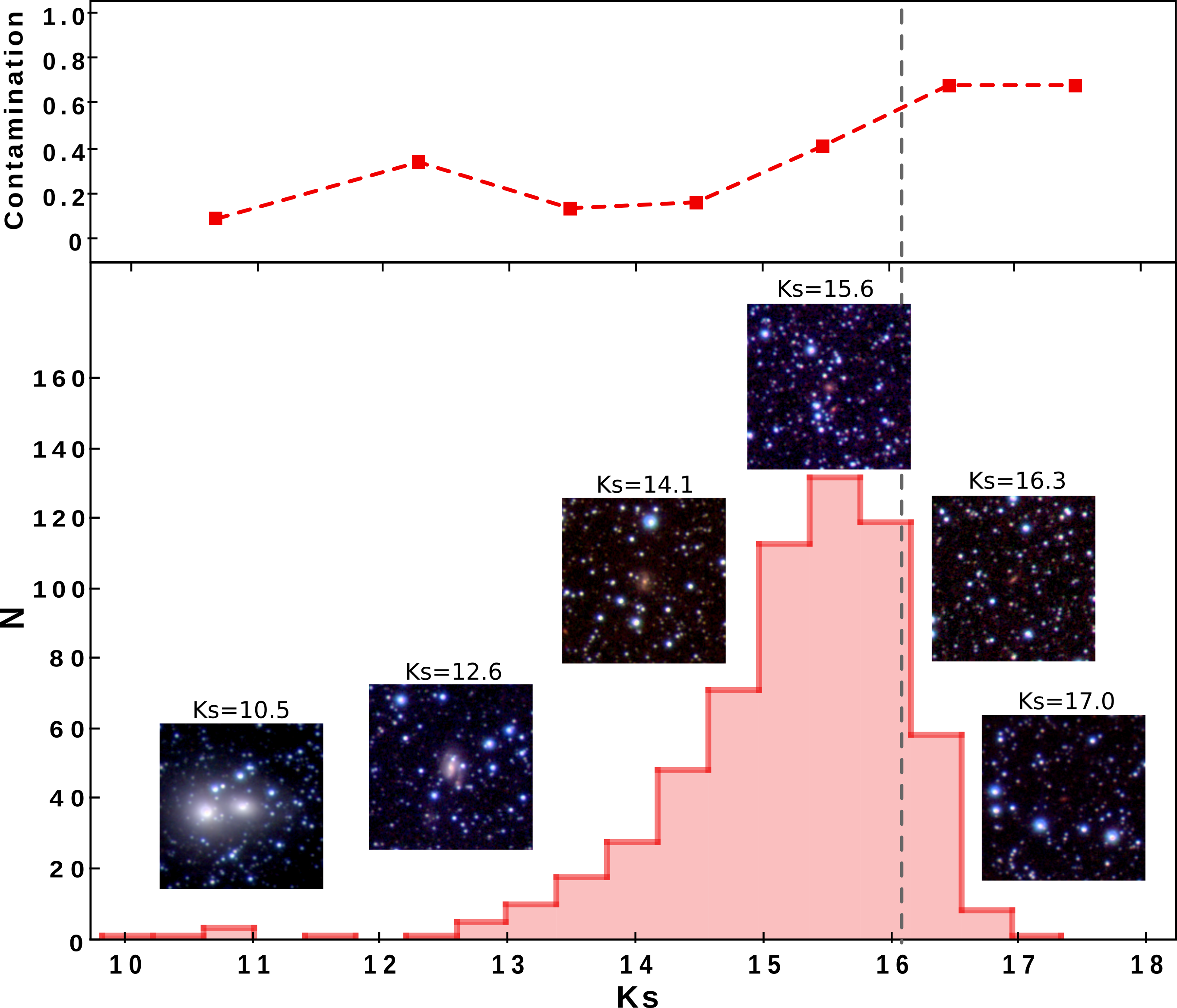}
\caption{Top: Contamination of the automatic catalogue of extended sources before visual inspection as a function of increasing Ks total magnitude. Bottom: Ks total magnitude distribution of visually confirmed galaxies and example images with different magnitudes. The vertical dashed line represents the reliability limit Ks$=$16.2 for visual inspection.} 
\label{KsdistCont}
\end{figure}

Given that the reliability of our catalogue grows with visual inspection, it is important to estimate how confident we are that these objects are real galaxies. To this end we explore galaxy images as a function of Ks magnitude. In Fig.~\ref{KsdistCont} (bottom panel) we plot Ks distribution of visually confirmed galaxies and example images with different magnitudes. From this analysis we find that our visual inspection is confident up to magnitude Ks=16.2. It is important to highlight that 90\% of visually confirmed galaxies are brighter than this limit.

 As a complementary study we calculated the contamination of the automatic catalogue of extended sources before visual inspection. In Fig.~\ref{KsdistCont} (top panel) we plot contamination as a function of increasing Ks total magnitude. From this figure, the contamination can be seen to grow toward fainter magnitudes and is higher than  50\% for galaxies with Ks$>$16.2. Therefore, visual inspection is a crucial step in the identification of galaxies using VVV images.

In addition, we analysed the reliability of the photometry of extended objects depending on the stellar density in the field.
The tile analysed in this work is located at the edge of the Galactic bulge, where the stellar density is lower than in the central region. We find no extended sources with a star at $d < 1 * r_{1/2}$ in the b204 tile. Furthermore, of a total of 624 objects identified as galaxy candidates, only five have a star closer than $d = 3 * r_{1/2}$. These five objects are also galaxies catalogued by 2MASX. Subsequently, considering both VVV and 2MASX, we calculated the magnitude differences in J, H, and Ks filters for these five galaxy candidates following \citet{GonzalezFernandez2018}.
We obtain good agreement in magnitudes between both catalogues, with an average difference of around 0.13 mag considering J, H, and Ks filters.

\subsection{Galaxy colours}
\label{cc}

\begin{figure}[htb]
\centering
\includegraphics[width=\columnwidth ]{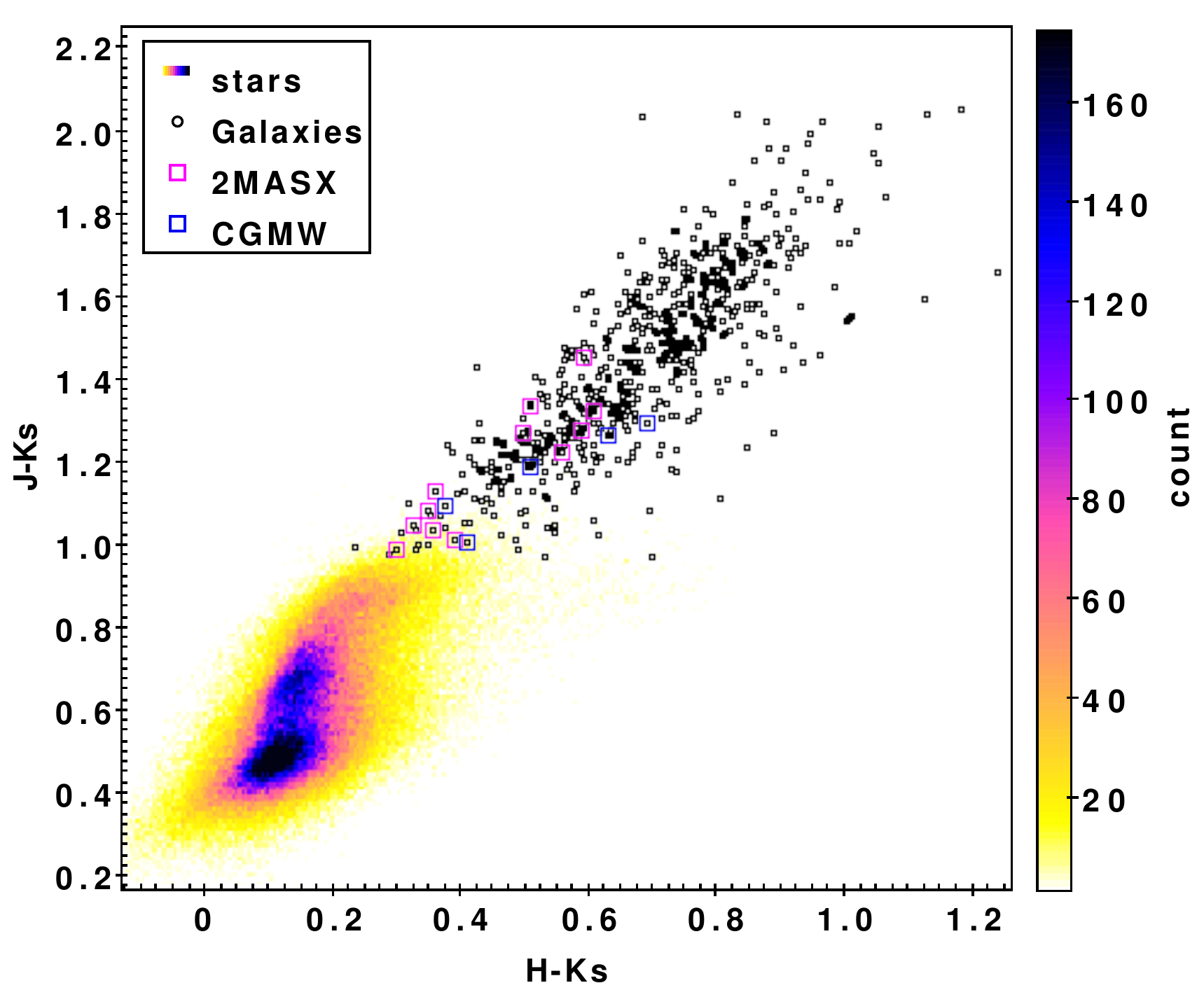}
\caption{J-Ks vs. H-Ks diagram.  The circles correspond to galaxy candidates. Galaxies catalogued by other surveys are shown in squares. The density map shows the distribution of stars and colour scale corresponds to number counts, as shown in the key.}
\label{colcol}
\end{figure}

Galaxies and stars have different colour distributions, and therefore we analysed photometric relations of galaxy candidates using the colour-colour and colour-magnitude diagrams. Figure~\ref{colcol} shows the J-Ks versus H-Ks diagram. 
A remarkable difference between stellar objects and extended sources is observed, the galaxy candidates being the farthest away from the whole sample. These results are in agreement with those of \cite{amo12} and \cite{col14}; in both cases, we observe a significant separation of galaxy candidates from the stars sample.

\begin{figure}[htb]
\centering
\includegraphics[width=\columnwidth]{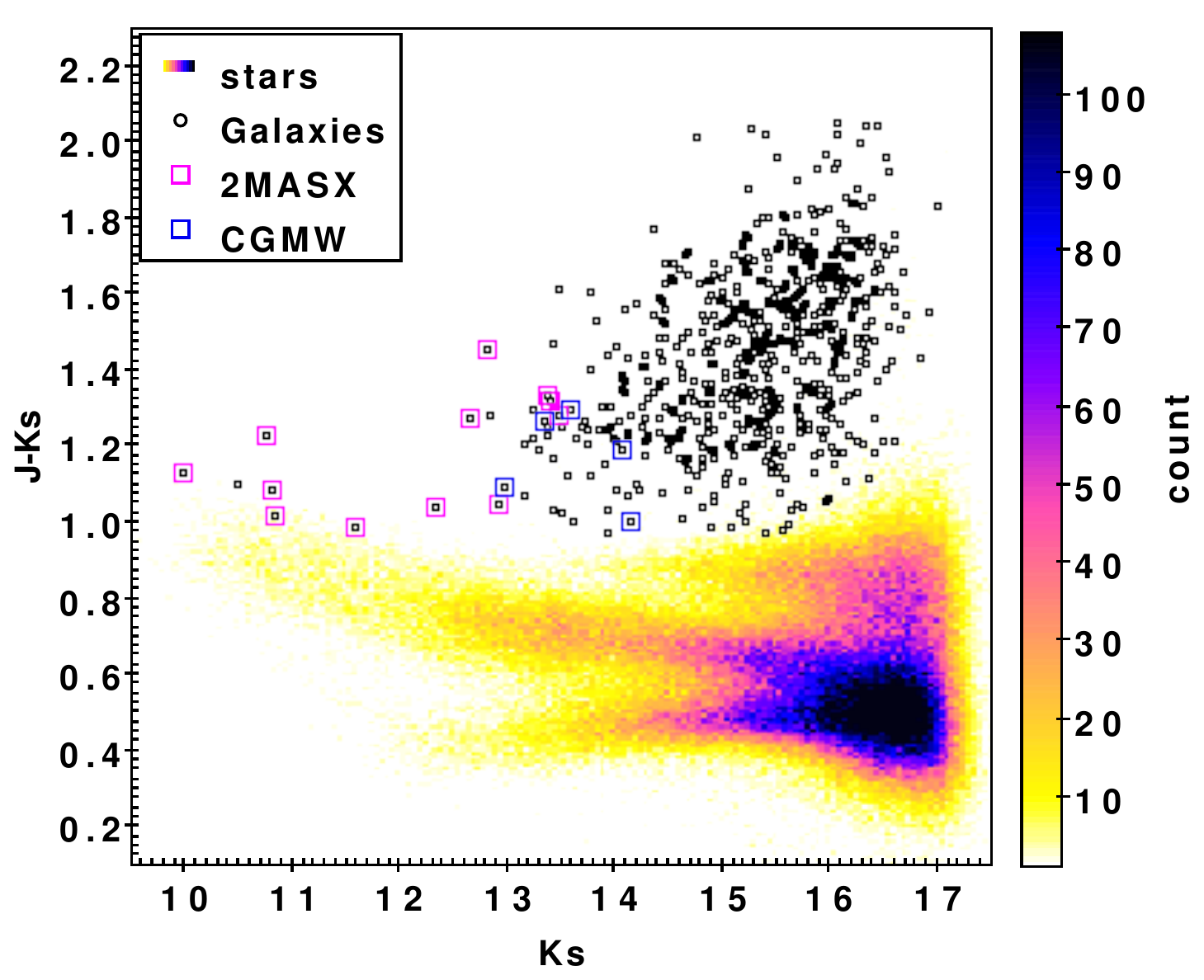}
\caption{J-Ks vs. Ks total magnitude diagram. Circles correspond to galaxy candidates. Galaxies catalogued by other surveys are shown by squares. The density map shows the distribution of stars and the colour scale corresponds to number counts, as shown in the key.}

\label{colmag}
\end{figure}

Figure~\ref{colmag} shows the J$-$Ks colour$-$magnitude diagram. We observe that the galaxy candidates occupy a particular position consistent with the results from \cite{amo12} and \cite{col14}. Most galaxies lie above the threshold  J-Ks$=$0.97  with a mean colour of 1.45 and a dispersion of 0.23. Galaxies catalogued by other surveys are in general brighter than the remaining galaxy candidates which demonstrates the potential of VVV data to identify fainter extended sources.

\section{Mock catalogues}
\label{sec:mock}

In order to estimate the total number of background galaxies to be found in the VVV bulge region, we constructed mock galaxy catalogues from two different SAMs of galaxy formation: L-Galaxies and SAG.

L-Galaxies \citep{Guo2011, Henriques2012} is applied to the dark matter subhaloes identified in the Millennium N-body simulations \citep{Millenium2005, Lemson2006}. 
Millennium cosmological simulations follows the evolution of $2160^{3}$ dark matter particles with masses of $8.6 \times 10^8 \,h^{-1} M_{\odot}$ in a comoving box of $500 \,h^{-1} \mathrm{Mpc}$ on a side, adopting a flat $\Lambda$CDM cosmological model.

On the other hand, a Semi-Analytic Galaxies (SAG) model has been developed from the version of \citet{springel2001}, and improved as described in \citet{Cora2006}, \citet{Lagos2008}, \citet{Tecce2010}, \citet{Orsi2014}, \cite{Padilla2014}, \citet{rui15} and \cite{Gargiulo2015}. The catalogue used in this work was constructed applying the latest version of SAG \citep{Cora2018, Collacchioni2018, Cora2019} to the Small MultiDark Planck simulation\footnote{doi:10.17876/cosmosim/smdpl} \citep[SMDPL,][]{Klypin2016}. The SMDPL simulation follows the evolution of $3840^{3}$ dark matter particles within a box of side-length $400 \,h^{-1} \mathrm{Mpc}$, resulting in a mass resolution of $m_{\rm p} = 9.63 \times 10^7\, h^{-1}$M$_{\odot}$ per particle. 
The cosmological parameters adopted in this simulation correspond to a flat $\Lambda$CDM model consistent with the results of \citet{Planck2014}.

Both semi-analytic galaxy catalogues include SDSS magnitudes, star formation histories, stellar masses of the disk and bulge components, dark matter host halo masses, and galaxy type, among other properties. 

To construct the mock catalogues, we first transformed the SDSS magnitudes to 2MASS magnitudes using the relations presented in \citet{bilir2008}, and then from 2MASS to VVV magnitudes with the relations described in \citet{GonzalezFernandez2018}. Afterwards, we placed a virtual observer in a corner of the simulation box at $z=0$ and applied an angular mask of the size of a VVV tile: $1.475 \times1.109$ deg$^2$. For each galaxy, we computed an observational redshift given by 

\begin{equation}
    z_{\rm obs} =  z_{\rm cos} + \frac{v_{\rm pec}}{c} (1 + z_{\rm cos}),
	\label{eq:observated_redshift}
\end{equation}

\noindent where $z_{\rm cos}$ is the cosmological redshift computed using the distance between the galaxy and the virtual observer, $v_{\rm pec}$ is the peculiar velocity of the galaxy in the direction of the line of sight, and $c$ is the speed of light.

Taking into account that our mocks cover a range in redshift up to $z=0.2$, we applied inverse K-corrections to the apparent magnitudes following the model of \citet{Chilingarian2010}. 
The next step is to compute apparent magnitudes from the absolute magnitudes provided by both SAMs. In addition to the dust extinction models implemented by each SAM to correct their magnitudes, our mock catalogues seek to simulate a very specific region of the sky, where the line of sight points to the Milky Way bulge, and therefore the dust absorption is higher than usual. For this reason,  two  extinction  models  were  adopted, namely those of \citet{Schlafly2011} and \citet{gon12}, in  order to obtain different estimations. In addition, for the sake of consistency, we applied cuts in apparent magnitude $10< \mathrm{Ks} < 17$ and colour $\mathrm{J-Ks} > 0.97$, $\mathrm{J-H} > 0$ and $\mathrm{H-Ks}>0$ to limit the mock catalogue galaxies to have similar photometric properties to the observational sample of galaxy candidates.
We found that more than 95\% of the galaxies from the mock catalogues fit in the adopted ranges used to define VVV galaxy candidates.

Due to the fact that our mock catalogues cover a small area in the sky, the cosmic variance must be taken into account. For each simulation, we performed 190 galaxy mocks, changing the line of sight direction, and then applied the angular mask of the VVV.
As we do not have spectroscopic information for the galaxy candidates, for each of these mocks, we computed and reported the integrated mean number of galaxies we find up to the redshift  limit used to construct the light cones. This procedure allows us to take into account the cosmic variance of the simulations when we consider small volume samples such as the one we used in our analysis.

\begin{figure}[htb]
	\includegraphics[width=\columnwidth]{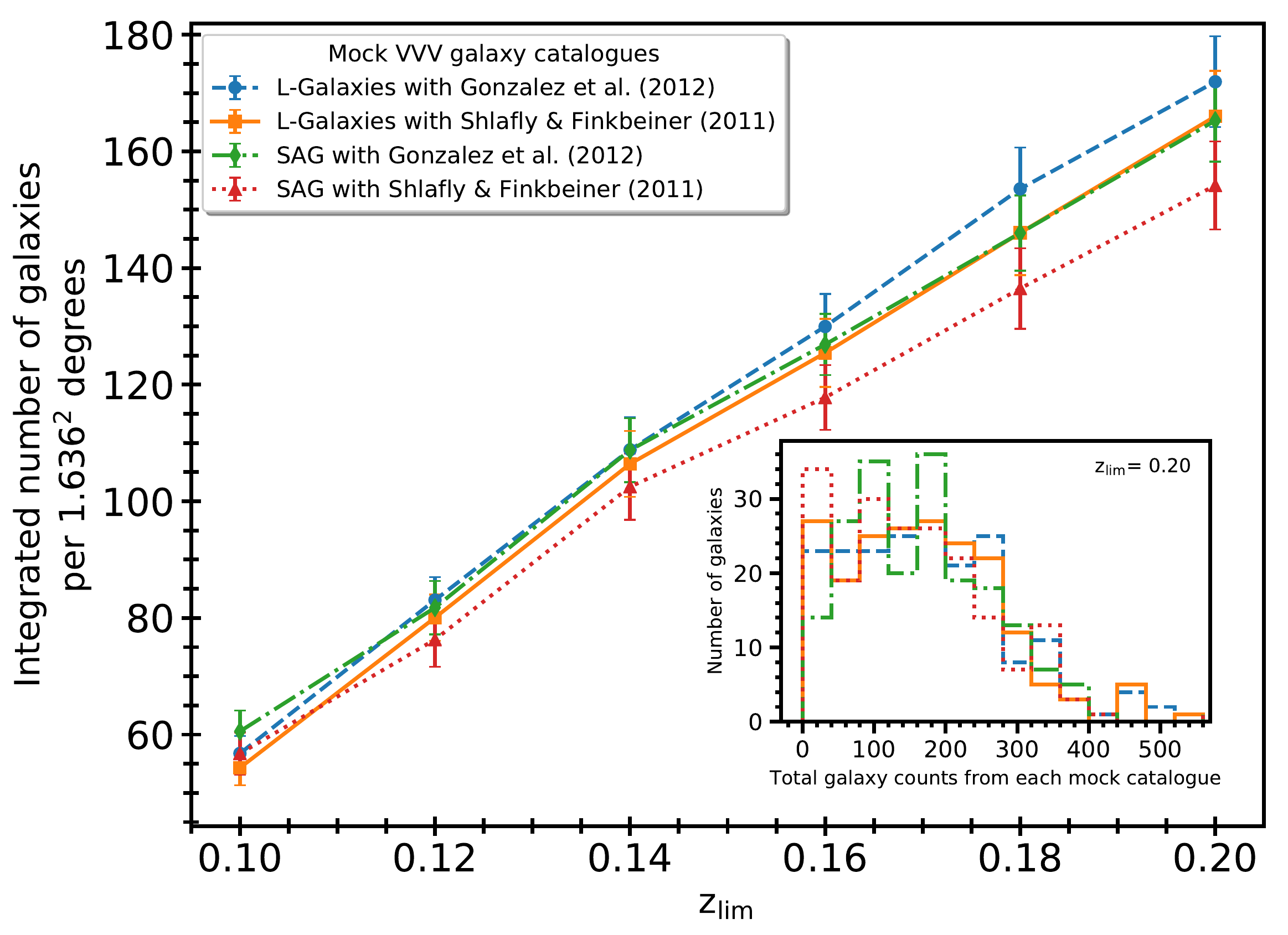}
    \caption{
    Integrated number of galaxies per $1.636\ \mathrm{deg}^{2}$ as a function of the redshift limits used to construct the light cones. The results for both SAG and L-Galaxies are shown, using the extinction maps of \citet{gon12} and \citet{Schlafly2011}, as indicated in the key. 
    The inset shows the total galaxy number counts from each mock catalogue by changing the line-of-sight direction of the light cone with a redshift limit of $0.20$.
    }
    
    \label{fig:number_vs_z}
\end{figure}

Figure~\ref{fig:number_vs_z} shows the integrated number of background galaxies per $1.636\ \mathrm{deg}^{2}$ as 
a function of the light cone redshift limit. As mentioned above, for both simulated catalogues we considered all galaxies with ${\rm Ks} < 17$ and two bulge extinction maps, as indicated in the key of the figure. In all cases, error bars were computed using the jackknife method.
Although the two synthetic galaxy catalogues were built with different SAMs applied on two different numerical cosmological simulations, the results are statistically consistent. 

On the other hand, the  observed differences, within the error bars, between the two extinction models can be explained from their mean extinctions.
For $b204$ VVV tile, \citet{gon12} have find an average extinction of $0.0018$ in $\mathrm{Ks}$ magnitude  while \citet{Schlafly2011} find a higher average value of $0.037$.
The inset panel in Fig.~\ref{fig:number_vs_z} shows the histograms of the total counts of galaxies with apparent magnitude $\mathrm{Ks} < 17$ where it is possible to observe the similarity between the mock catalogues with both extinction maps.

Considering the mock catalogues for background galaxies with the two extinction maps, we obtain an integrated mean value of approximately $60$ and $160$ galaxies per tile for the limit redshift $z_{lim} = 0.10$ and $z_{lim} = 0.20$, respectively. 
For the sake of clarity, in Table \ref{tab3} we also show the integrated average number of galaxies behind the bulge within $1\ \mathrm{deg}^{2}$.

Furthermore, we performed the counts by directing the line of sight of the mock catalogues towards an overdense region in the simulation in order to estimate the galaxy counts in an area of enhanced density. The values are also reported in Table \ref{tab3}. In addition, even though redshift information is not available for galaxies in our sample, the measured photometric properties of these candidates are consistent with an average redshift within the analysed range, in agreement with the work of \cite{col14}.

Notably, in this study we found more than 600 galaxy candidates in the $b204$ VVV tile, which is clear independent evidence of galaxy overdensity if we consider the expected integrated number of galaxies from the lowest or highest redshift used to build the mock catalogues, that is $z_{lim} = 0.10$ or $z_{lim}  = 0.20$, respectively. It is important to note that this finding is consistent with the number of galaxies per $1\ \mathrm{deg}^{2}$ obtained when the light cones points towards overdense regions (see Table \ref{tab3}).
This result may indicate the presence of a galaxy cluster in the $b204$ VVV tile.

\begin{figure}[htb]
    \centering
    \includegraphics[width=\columnwidth]{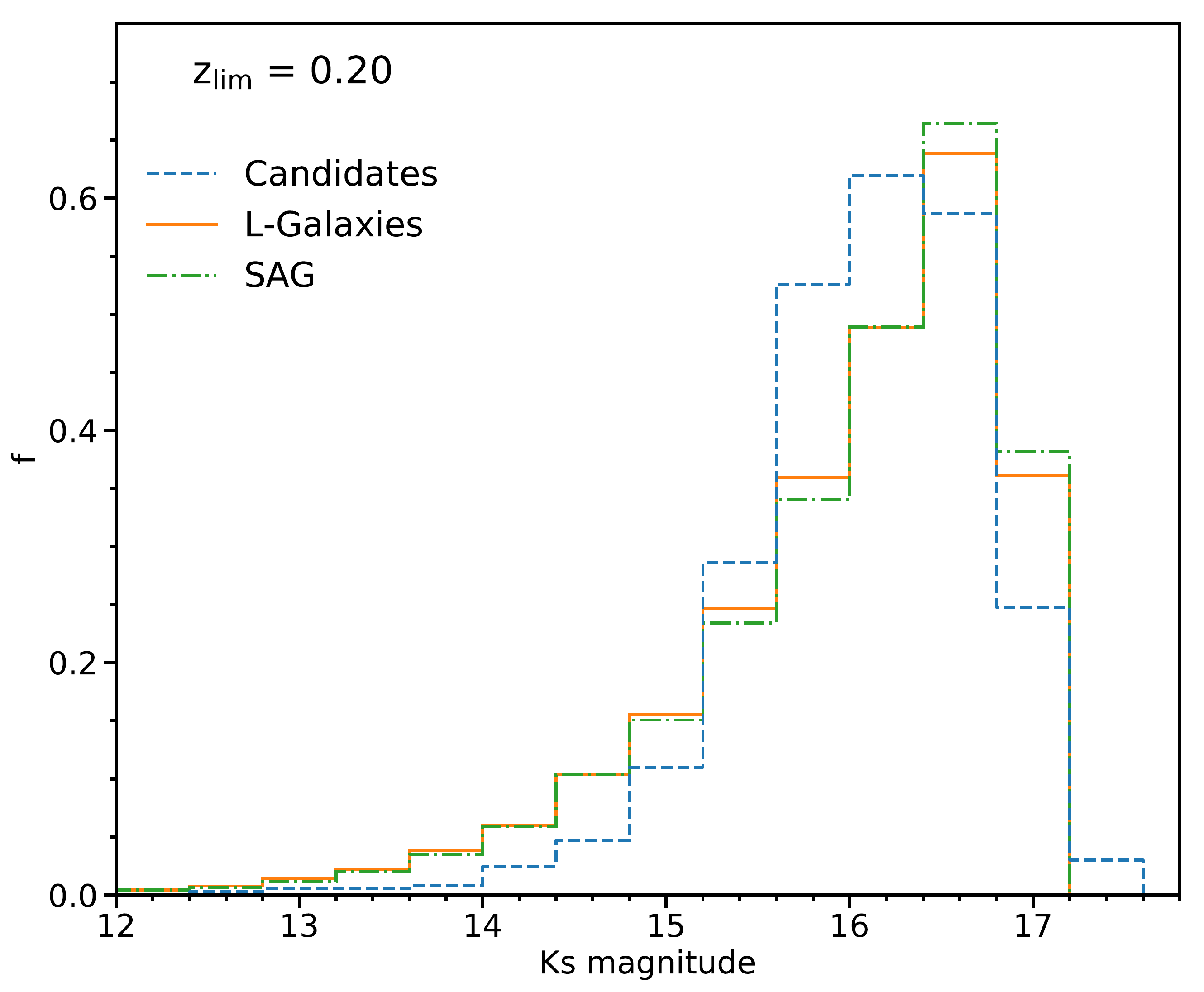}
	   \caption{Normalised distributions of Ks magnitude of galaxy candidates in tile $b204$ (blue dashed line), compared with those from L-Galaxies and SAG with the \citet{gon12} extinction model (continuous lines of orange and green colours, respectively). 
	   }
    \label{fig:Ks_distributions}
\end{figure}

In order to compare the observational data with those derived from the Mock catalogue, the normalised distributions of Ks magnitude for the SAMs
with the \citet{gon12} extinction model and the galaxy candidates at redshift limit $\mathrm{z} = 0.20$ are plotted in Fig.~\ref{fig:Ks_distributions} (the two extinction models show similar behaviour).
We find reasonable agreement between the magnitude distributions of both models, although there are differences in the distribution of the observed galaxy candidates. 
In addition, the average values of $\mathrm{J-Ks}$ for both SAMs ($1.43 \pm 0.51$ and  $1.46 \pm 0.52$ for L-Galaxies and SAG with \citet{gon12}, respectively) are in good agreement with the observational data supporting the results. The observed discrepancies can be traced back to the choices made in the SAMs (e.g. the stellar population synthesis model used to build up the galaxy luminosity, and the evolution of the properties of galaxies with time, which may be affected by the physical recipes used in these models).

\begin{center}
\begin{table*}
\caption{
Integrated number of galaxies per $1~{\rm deg}^2$ for different redshift limits toward the mean and overdense line-of-sight directions.
} 
\vskip 0.1cm
\begin{flushright}
\begin{tabular}{|c|cc|cc|cc|cc|}
\hline
&\multicolumn{2}{|c|} {} &\multicolumn{2}{|c|} {}&\multicolumn{2}{|c|} {}&\multicolumn{2}{|c|} {}\\
&\multicolumn{2}{|c|} {
\begin{tabular}{@{}c@{}} 
L-Galaxies\\ with Gonzalez\\ et al. (2012)
\end{tabular}} 
&\multicolumn{2}{|c|} {
\begin{tabular}{@{}c@{}} 
L-Galaxies\\ with Schlafly\\ \& Finkbeiner\\ (2011)
\end{tabular}} 
&\multicolumn{2}{|c|} {
\begin{tabular}{@{}c@{}} 
SAG with\\ Gonzalez\\ et al. (2012)
\end{tabular}}
&\multicolumn{2}{|c|} {
\begin{tabular}{@{}c@{}} 
SAG with\\ Schlafly \&\\ Finkbeiner\\ (2011)
\end{tabular}}\\
&\multicolumn{2}{|c|} {} &\multicolumn{2}{|c|} {}&\multicolumn{2}{|c|} {}&\multicolumn{2}{|c|} {}\\
\hline
\hline
\textit{z}
&\begin{tabular}{@{}c@{}} 
Mean\\ background\\ region
\end{tabular}
&\begin{tabular}{@{}c@{}} 
Overdense\\ background\\ region
\end{tabular}
&\begin{tabular}{@{}c@{}} 
Mean\\ background\\ region
\end{tabular}
&\begin{tabular}{@{}c@{}} 
Overdense\\ background\\ region
\end{tabular}
&\begin{tabular}{@{}c@{}} 
Mean\\ background\\ region
\end{tabular}
&\begin{tabular}{@{}c@{}} 
Overdense\\ background\\ region
\end{tabular}
&\begin{tabular}{@{}c@{}} 
Mean\\ background\\ region
\end{tabular}
&\begin{tabular}{@{}c@{}} 
Overdense\\ background\\ region
\end{tabular}\\
\hline
0.10 & 34.5 $\pm$ 1.8 & 152.3 $\pm$ 4.0 & 33.2 $\pm$ 1.8 & 148.2 $\pm$ 4.4 & 37.2 $\pm$ 2.2 & 157.3 $\pm$ 3.7 & 34.9 $\pm$ 2.2 & 147.0 $\pm$ 4.6\\ 
0.12 & 50.5 $\pm$ 2.3 & 161.8 $\pm$ 4.4 & 48.8 $\pm$ 2.4 & 156.9 $\pm$ 4.8 & 49.8 $\pm$ 2.9 & 149.2 $\pm$ 4.0 & 46.3 $\pm$ 2.8 & 138.7 $\pm$ 4.7\\
0.14 & 66.6 $\pm$ 3.4 & 193.9 $\pm$ 5.6 & 65.0 $\pm$ 3.5 & 183.3 $\pm$ 5.9 & 67.6 $\pm$ 3.4 & 216.1 $\pm$ 5.6 & 63.7 $\pm$ 3.6 & 202.0 $\pm$ 6.7\\
0.16 & 78.8 $\pm$ 3.4 & 219.5 $\pm$ 6.5 & 76.1 $\pm$ 3.5 & 213.4 $\pm$ 7.0 & 77.6 $\pm$ 3.2 & 340.1 $\pm$ 8.5 & 72.1 $\pm$ 3.4 & 317.5 $\pm$ 10.4\\ 
0.18 & 92.6 $\pm$ 4.3 & 321.6 $\pm$  9.5 & 88.3 $\pm$ 4.4 & 315.9 $\pm$ 10.3 & 88.4 $\pm$ 4.0 & 378.5 $\pm$  9.7 & 82.7 $\pm$ 4.2 & 352.5 $\pm$ 11.8\\ 
0.20 & 105.3 $\pm$ 4.8 & 373.3 $\pm$ 11.4 & 101.5 $\pm$ 4.8 & 358.0 $\pm$ 12.1 & 101.3 $\pm$ 4.4 & 476.2 $\pm$ 11.6 & 94.8 $\pm$ 4.6 & 444.2 $\pm$ 14.6\\
\hline
\hline
\end{tabular}
\label{tab3}
\end{flushright}
\end{table*}
\end{center}

\section{Galaxy overdensity}
\label{sec:Overdensity}
\begin{figure}[htb]
\centering
\includegraphics[width=0.5\textwidth ]{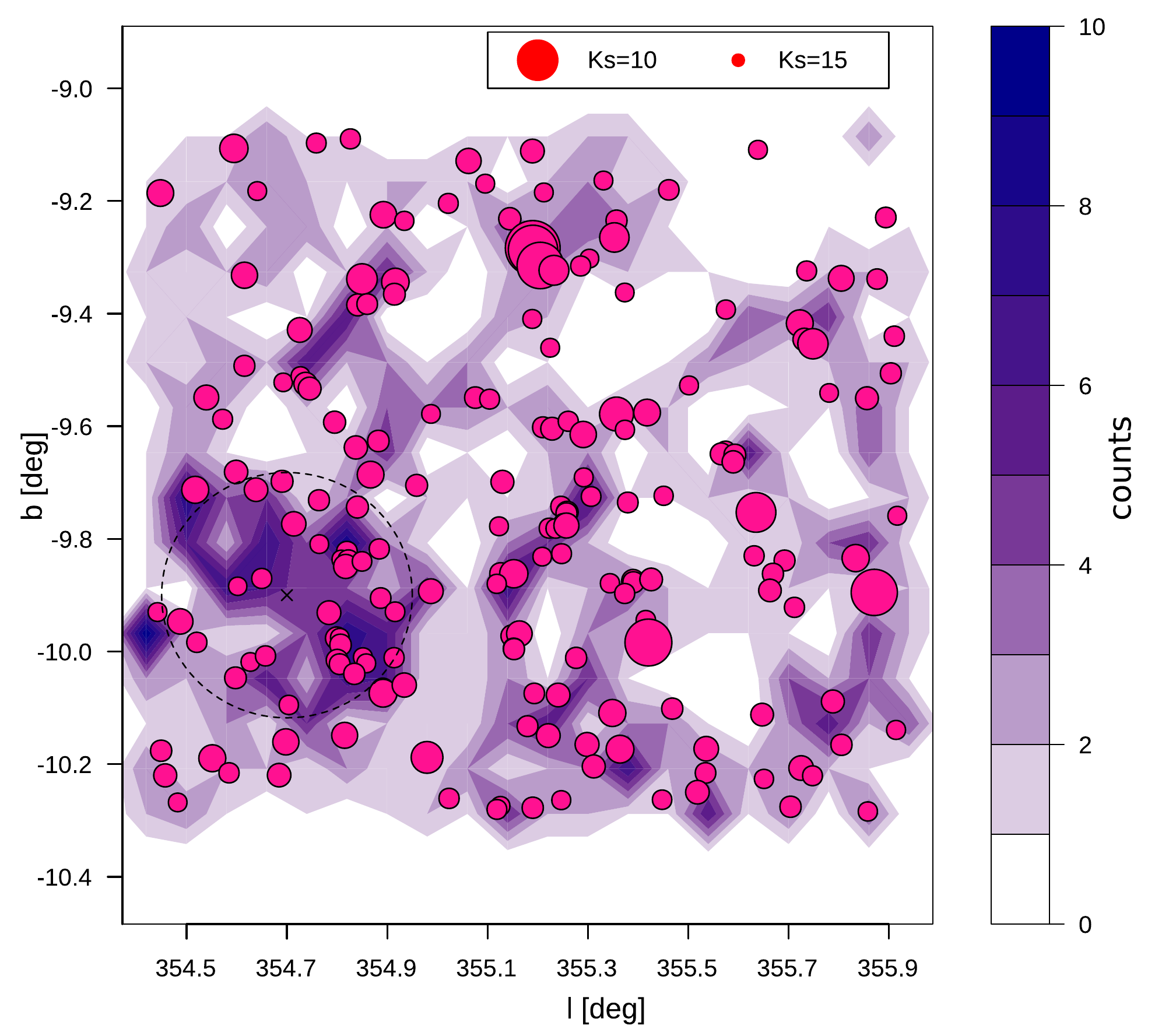}
\caption{Density map of visually confirmed galaxies. Filled contours are colour$-$codded according to number counts in pixels of 4 arcmin x 4 arcmin. Dots indicate the positions of the 30\% brightest galaxies (dot sizes are weighted according size is proportional to Ks magnitude). The dashed circle, of 15 arcmin radius is centred at the position of the overdense region in the tile.}
\label{radec}
\end{figure}

We constructed a density map of visually confirmed galaxies to reveal the distribution of the sources in the plane of the sky (Fig.~\ref{radec}). We find some regions of high concentration. In particular, the zone with major overdensity close to coordinates $l=354.66^o$ and $b=-9.85^o$ ($\alpha=273.6^o$ and $\delta=-38.4^o$ (J2000)) occupies an approximate area with a radius of 15 arcmin (the circle in Fig.~\ref{radec}). This area represents approximately 12\% of the tile and contains at least 19\% ($\approx 118$) of the visual galaxies in our catalogue.

In order to validate this finding, we calculated the number of galaxies in 20 random circular regions of 15 arcmin radius within the tile. On average, each one of these random regions contains only 6.7\%($\approx 42$) of the galaxy candidates. This result shows that the region of major overdensity is approximately three times denser than the randomly selected areas, confirming that this latter region contains a genuine overdensity of galaxies.

\begin{figure}[htb]
\centering
\includegraphics[width=\columnwidth]{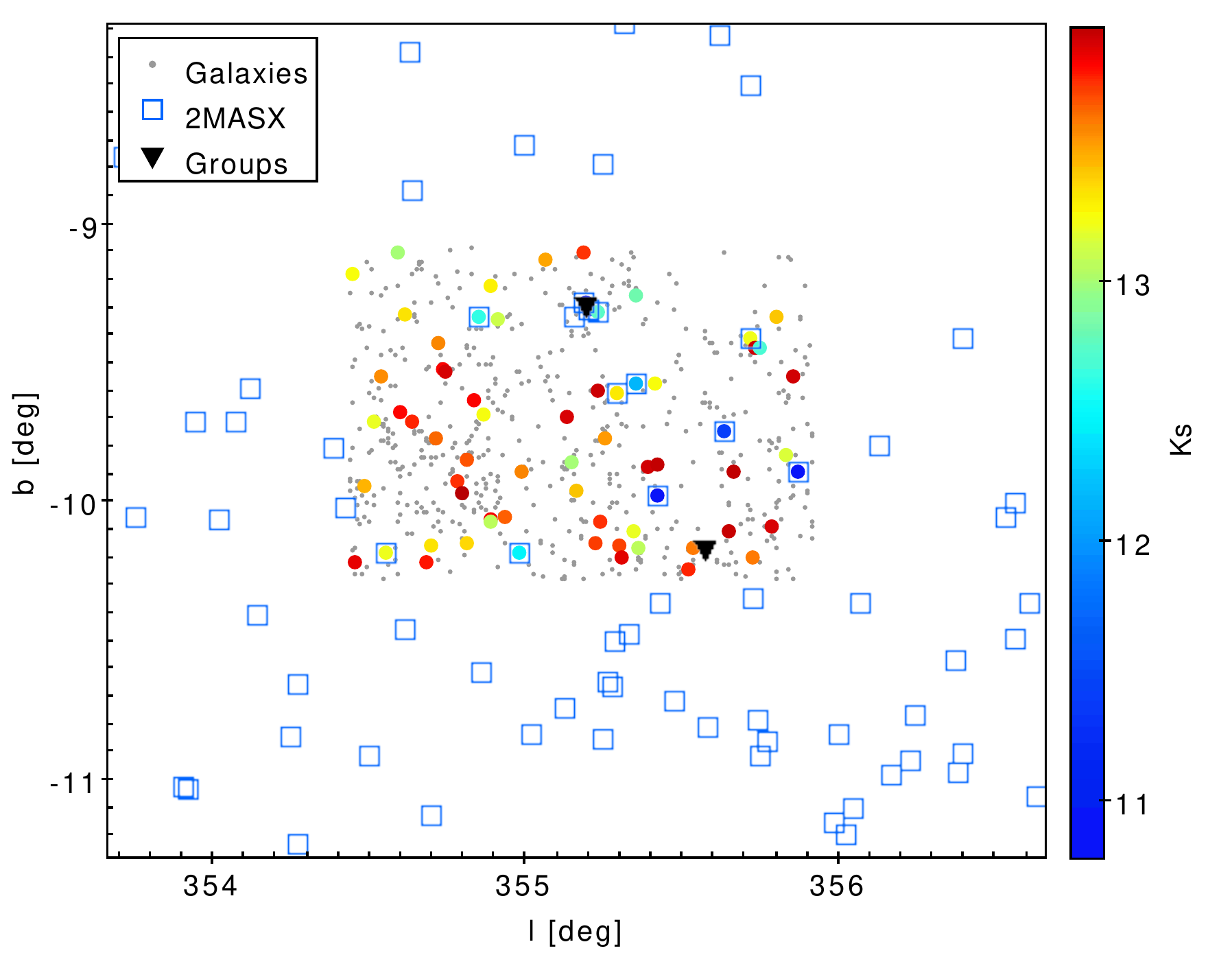}
\caption{Distribution of 2MASX galaxies around tile $b204$ (open squares). Galaxies are shown as grey dots and 10\% of the brightest visually confirmed galaxies are colour-coded according to Ks magnitude. The filled triangles represent the position of \cite{Tempel2018} 2MRS groups. 
}
\label{gxgroups}
\end{figure}

We also explored the distribution of 2MASX galaxies from  \citet{Jarrett2000} and galaxy groups from \citet{Tempel2018} around  tile $b204$ (Fig. \ref{gxgroups}). The sample of galaxies is complete up to magnitude Ks$=13.5$ and galaxy groups are detected from the 2MRS data set described in \citet{Huchra2012}, which includes galaxies
brighter than Ks$=11.75$. We find in Fig. \ref{gxgroups} that number counts of 2MASX galaxies seem to increase towards higher Galactic latitudes, finding more galaxies at $b<-10$. This is in agreement with results from \citet{Jarrett2000b}, who found a 30\% $-$ 50\% decline in source counts for Galactic latitudes between 5 and 12 degrees, as a consequence of an increment in stellar density that affects the identification of extragalactic objects. Only two galaxy groups are catalogued in the the area studied, both with only two galaxy members and none of them associated with the overdensity described above.

On the other hand, \cite{amo12} catalogued 214 new galaxy candidates in a tile area near the Galactic plane using the VVV data in the disk tile $d003$ centred at $l=298.356^o$ and $b=-1.650^o$ where the extinction is also relatively low and uniform, reaching a mean value of $A_{Ks} = 0.549$, in agreement with the expected one by Millennium simulations used in their analysis. 
These latter authors identified 72 objects as Type I sources (possible galaxies) and 142 as Type II (likely galaxies).
In addition, \cite{col14} examined the VVV $b261$ tile centred at $l=356.597^o$ and $b=-5.321^o$  with NIR photometry of VVV survey data. The mean values of extinction were $A_{J} = 0.26$, $A_{H} = 0.14$, and $A_{Ks} = 0.09$. The authors detected at least fifteen galaxies in the central area (350 kpc from the X-ray peak emission) of the X$-$ray$-$detected galaxy cluster Suzaku J1759-3450, at z = 0.13.
In a recent study, \cite{bar18} studied two disk tiles, $d010$ centred at $l=308.569^o$ and $b=-1.650^o$ and $d115$ centred at $l=295.438^o$ and $b=1.627^o$. The obtained median $A_{Ks}$ values were $0.86 \pm 0.32$ and $0.42 \pm 0.08$ for the d010 and d115 tiles, respectively. The authors found 345 and 185 extragalactic candidates in $d010$ and $d115$, respectively, making a total of 530 sources detected with the VVV survey in two regions of the disk.

With the present study, we reveal more than 600 galaxy candidates, a higher number than in previous works. Nevertheless the extinction  for b204 tile is lower than for the disk tiles d003, d010, and d115, and therefore galaxy number counts in $b204$ could be similar to the disk ones at comparable extinction levels. 
However, it is worth noting that we identify a high concentration of galaxies in a small region of 15 arcmin radius with three times higher density than the remaining area, even though the whole tile has a uniform extinction. Also, galaxy number counts are approximately three times the amount estimated by our mock catalogues for a mean background region despite the fact that the star density is much higher in a bulge tile than in a disk tile which may be detrimentally affecting our search for galaxies.
Furthermore, the density in tile $b204$ is consistent with the mock counts calculated by directing the line of sight towards an overdense region in the simulation. Therefore, we speculate on the existence of an overdensity of galaxies, considering  this finding as a possible cluster.

\section{Summary}
\label{sec:conc}

The Milky Way represents an obstacle for observations at optical wavelength of extragalactic sources. For this reason clear information about the galaxies and large extragalactic structures in this region is lacking. To obtain information for this area, we used the VVV survey data and analysed the tile $b204$ located in the central region of the Milky Way. 

We setup an automatic detection procedure using the software SExtractor. In this way, we generated a multi-band catalogue with stellar and extragalactic sources. This sample was calibrated with the CASU catalogue in order to improve photometric and astrometric precision. This step was carried out using bright but non$-$saturated stars. 

In order to obtain a catalogue of automatically selected extended sources we used the star classification parameter, CLASS\_STAR, and the half$-$light radius, $r_{1/2}$, given by SExtractor.
In this sense we considered objects with CLASS\_STAR$< 0.5$ and $r_{1/2} > 0.7$ arcsec as likely extended.
In addition, to separate galaxy candidates from the stellar locus we selected objects with Ks$>$10 and considered specific colour cuts (J-Ks$>$0.97, J-H$>$0 and H-Ks$>$0), following previous works \citep {amo12,col14,bar18}.
We also performed a visual inspection in false$-$colour RGB images  constructed from filters J, H and Ks. We select as visual galaxy candidates those objects showing morphological features and surface brightness corresponding to galaxies.
Following this procedure, we obtained 624 galaxy candidates behind the Galactic bulge.

To test the reliability of our methods we used  13 2MASX and five CGMW galaxies catalogued in tile $b204$. Because most catalogued objects are 2MASX galaxies we restricted our analysis to the completeness limit of this sample, 10$<$Ks$<$13.5. Subsequently, to select colour cuts we evaluated different values based on the colour of the 2MASX sources and Mock galaxies. We found J-H$>$0, H-Ks$>$0, and J-Ks$>$0.97 to be a good compromise between completeness and contamination. If we consider only catalogued sources as reliable galaxies (an unrealistic assumption given that the VVV images are much deeper) the contamination of the extended sources catalogue automatically identified by SExtractor is 60\%; but when taking visually confirmed galaxies as reliable objects the contamination drops to 28\%. Also, with the adopted colour cuts we lose only one 2MASX catalogue source and one extended object that is visually confirmed as a galaxy, and therefore the completeness of our visual catalogue isestimated as 87\% up to magnitude Ks=13.5.
Given the fact that the reliability of our catalogue grows with visual inspection, we studied contamination of the automatic catalogue of extended sources before visual inspection as a function of Ks magnitude, finding a reliability limit of Ks=16.2. It is important to highlight that 90\% of visually confirmed galaxies are brighter than this limit.

Using J-Ks versus H-Ks colour$-$colour and J-Ks colour$-$magnitude diagrams, we verified that the sample of galaxy candidates are separated from the general distribution of the stellar sources presenting clearly redder colours.

We analysed the spatial distribution of galaxy candidates, finding an overdense region. We constructed two mock background galaxy catalogues and considered two extinction maps. 
The values obtained from the mock catalogues by directing the light cone towards an overdensity region are in good agreement with the values found in this work.
These results provide clear evidence of a galaxy overdensity in the $b204$ VVV tile.
We have therefore requested follow-up telescope time to obtain the spectra of the brightest galaxies in these overdensity region. The spectroscopic data will allow us to measure redshifts and confirm the nature of this structure in the near future.

In total, we find 624 extended sources (including 607 new galaxy candidates that have been catalogued for the first time
and 17 galaxies catalogued by other surveys). In addition, we did not identify any X-ray or radio sources in this region. Moreover, our study is the first of its kind to cover the bulge region, whereby a systematic search for extragalactic sources was performed using the VVV survey. We wish to highlight the fact that the methods used in this work are a starting point in the development of techniques devoted to identifying extragalactic sources, such as galaxies, groups, and clusters, in the bulge area mapped by the VVV survey. Also, in future works we will use the VVVX survey \citep{2016gsnr.confE..10M}, which will cover a larger area (three times that of the VVV), making it possible to search for more structures behind the bulge.

It is worth noting that by using the VVV NIR images, we were able to detect new extragalactic sources that have not been detected with other surveys. Moreover, we implemented mock catalogues to statistically define an overdensity region probably corresponding to a galaxy system like a group or galaxy cluster. Therefore, this paper and previous works demonstrate the potentiality of the VVV survey to find and study a large number of galaxy candidates and extragalactic structures obscured by the Milky Way.

\begin{acknowledgements}
We thank the referee for providing us with helpful comments that improved this paper. This work was partially supported by the Consejo Nacional de Investigaciones Cient\'{\i}ficas y T\'ecnicas and the Secretar\'{\i}a de Ciencia y T\'ecnica de la Universidad Nacional de San Juan. 
The authors gratefully acknowledge data from the ESO Public Survey program ID 179.B-2002 taken with the VISTA telescope, and products from the Cambridge Astronomical Survey Unit (CASU). D.M. acknowledges support from the BASAL Center for Astrophysics and Associated Technologies (CATA) through grant PFB-06, the Ministry for the Economy, Development and Tourism, Programa Iniciativa Cient\'ifica Milenio grant IC120009, awarded to the Millennium Institute of Astrophysics (MAS), and the FONDECYT Regular grants No. 1170121.
CVM acknowledges financial support from the Max Planck Society through a Partner Group grant. The authors gratefully acknowledge the Gauss Centre for Supercomputing e.V.
(www.gauss-centre.eu) and the Partnership for Advanced Supercomputing in Europe
(PRACE, www.prace-ri.eu) for funding the \textsc{MultiDark} simulation project by providing computing time on the GCS Supercomputer SuperMUC at Leibniz 
Supercomputing Centre (LRZ, www.lrz.de).
SAC acknowledges funding from {\it Consejo Nacional de Investigaciones Cient\'{\i}ficas y T\'ecnicas} (CONICET, PIP-0387), {\it Agencia Nacional de Promoci\'on Cient\'ifica y Tecnol\'ogica} (ANPCyT, PICT-2013-0317), and {\it Universidad Nacional de La Plata} (G11-150),Argentina.
ANR acknowledges funding from ANPCyT (PICT 2016-1975) and Secretar\'ia de Ciencia y Tecnolog\'ia, Universidad Nacional de C\'ordoba.  .
\end{acknowledgements}

\bibliography{references}

\end{document}